\documentclass{desyproc}
\usepackage{xspace}
\usepackage{relsize}
\usepackage{color}
\usepackage{graphicx}  % Include figure files
\usepackage{rotating}

\definecolor{Red}{rgb}{1,0,0}
\definecolor{Blue}{rgb}{0,0,1}
\definecolor{Green}{rgb}{0,1,0}

\newcommand\vud {\ensuremath{V_{\mathrm{ud}}}\xspace}
\newcommand\vus {\ensuremath{V_{\mathrm{us}}}\xspace}
\newcommand\vub {\ensuremath{V_{\mathrm{ub}}}\xspace}
\newcommand\vcd {\ensuremath{V_{\mathrm{cd}}}\xspace}
\newcommand\vcs {\ensuremath{V_{\mathrm{cs}}}\xspace}
\newcommand\vcb {\ensuremath{V_{\mathrm{cb}}}\xspace}
\newcommand\vtd {\ensuremath{V_{\mathrm{td}}}\xspace}
\newcommand\vts {\ensuremath{V_{\mathrm{ts}}}\xspace}
\newcommand\vtb {\ensuremath{V_{\mathrm{tb}}}\xspace}
\def\vckm       {\ensuremath{{V}_{\mathrm{CKM}}}\xspace}

\newcommand\modvub {\ensuremath{|V_{\mathrm{ub}}|}\xspace}

\newcommand\modvcb {\ensuremath{|V_{\mathrm{cb}}|}\xspace}
\newcommand\modvtd {\ensuremath{|V_{\mathrm{td}}|}\xspace}
\newcommand\modvts {\ensuremath{|V_{\mathrm{ts}}|}\xspace}

\def\theckmmatrix  {\ensuremath{ \left( \begin{array}{ccc} \vud & \vus & \vub \\ \vcd & \vcs & \vcb \\ \vtd & \vts & \vtb \end{array}\right)}}
\def\rhobar {\ensuremath{\overline{\rho}}\xspace}
\def\etabar {\ensuremath{\overline{\eta}}\xspace}

\def\sintwobeta{\ensuremath{\sin 2\beta}\xspace}
\def\sintwobetaeff{\ensuremath{\sin 2\beta_{\mathrm{eff}}}\xspace}
\def\betaeff{\ensuremath{\beta_{\mathrm{eff}}}\xspace}
\def\deltassm{\ensuremath{\Delta S_{\mathrm{SM}}}\xspace}
\def\deltas{\ensuremath{\Delta S}\xspace}
\def\ccbars{\ensuremath{c \overline{c} s}\xspace}

\def\alphaeff{\ensuremath{\alpha_{\mathrm{eff}}}\xspace}

\def\lambdacp{\ensuremath{\lambda_{\mathrm{CP}}}}

\def\qovp{\ensuremath \frac{q}{p}}
\def\modqovp{\ensuremath \left| \qovp \right|}

\def\dmd{\ensuremath{\Delta m_d}}

\def\BBb  {\ensuremath{B\overline{B}}\xspace}
\def\B    {\ensuremath{B}\xspace}
\def\Bz    {\ensuremath{B^{0}}\xspace}
\def\Bzb   {\ensuremath{\overline{B}^0}\xspace}
\def\BH    {\ensuremath{B_{\mathrm{H}}}\xspace}
\def\BL    {\ensuremath{B_{\mathrm{L}}}\xspace}
\def\BLH   {\ensuremath{B_{\mathrm{L,H}}}\xspace}
\def\Bp    {\ensuremath{B^+}\xspace}

\def\Bpm   {\ensuremath{B^\pm}\xspace}

\def\Bflav {\ensuremath{B_{\rm flav}}\xspace}
\def\Btag {\ensuremath{B_{\rm tag}}\xspace}
\def\Brec {\ensuremath{B_{\rm rec}}\xspace}
\def\Deltagamma {\ensuremath{\Delta\Gamma}\xspace}
\def\deltat {\ensuremath{\Delta t}\xspace}

\def\ks {\ensuremath{K^0_S}\xspace}

\def\kz {\ensuremath{K^0}\xspace}

\def\etap {\ensuremath{\eta^\prime}}

\def\C    {\ensuremath{C}\xspace}
\def\P    {\ensuremath{P}\xspace}
\def\T    {\ensuremath{T}\xspace}
\def\CP   {\ensuremath{CP}\xspace}
\def\CPT  {\ensuremath{CPT}\xspace}

\def\babar{\mbox{\slshape B\kern-0.1em{\smaller A}\kern-0.1em
    B\kern-0.1em{\smaller A\kern-0.2em R}}\xspace}
\def\belle{\mbox{\normalfont Belle}\xspace}

\def\jprBase{Phys.\ Rev.\xspace}
\def\jprBase{Phys.\ Rev.\xspace}
\newcommand\jprl[1] {\jprBase\ Lett. {\bf #1}}
\newcommand\jprd[1] {\jprBase\ D {\bf #1}}
\newcommand\jplb[1] {Phys.\ Lett.\ B {\bf #1}}
\newcommand\nima[1] {Nucl.\ Instrum.\ Meth.\ Phys.\ Res., Sec.\ A {\bf #1}}

\def\etal {{\em{at al.}}\xspace}
\def\epem {\ensuremath{e^+e^-}\xspace}

\def\ifb       {\ensuremath{\mathrm{fb}^{-1}}\xspace}
\def\iab       {\ensuremath{\mathrm{ab}^{-1}}\xspace}
\def\ps        {\ensuremath{\mathrm{ps}}\xspace}
\def\stat{\mathrm{(stat.)}}
\def\syst{\mathrm{(syst.)}}

\newcommand\ket[1]{\ensuremath{| #1 \rangle}\xspace}

\newcommand{\e}      [1]   { {\ensuremath{ \times 10^{ {#1} } }}}
\newcommand{\su}     [1]  {\ensuremath{SU(#1)}}

\begin{document}
%------------------------------------
\title{Results from the \B Factories}

%for single authors the superscripts are optional
\author{{\slshape Adrian Bevan$^1$}\\[1ex]
$^1$Department of Physics, Queen Mary, University of London, London E14NS, UK
}

% if the proceedings are available online (e.g. at Indico)
% please enter the contribution ID or file_name below for the DOI
%\contribID{32}
%\contribID{adrian\_bevan}

% TO THE CONFERENCE EDITORS:
% please update the following information
% before sending the template to the authors
% \confID{800}  % if the conference is on Indico uncomment this line
\desyproc{DESY-PROC-2008-xx}
\acronym{HQP08} % if you want the Acronym in the page footer uncomment this line
\doi  % if there is an online version we will register DOIs

\maketitle

\begin{abstract}
These proceedings are based on lectures given at the Helmholtz International Summer School Heavy Quark Physics at the
Bogoliubov Laboratory of Theoretical Physics, Dubna, Russia, during August 2008. I review the current status of CP
violation in \B meson decays from the \B factories.  These results can be used, along with measurements of the sides of
the Unitarity Triangle, to test the CKM mechanism.  In addition I discuss experimental studies of \B decays to final
states with `spin-one' particles.
\end{abstract}

%%%%%%%%%%%%%%%%%%%%%%%%%%%%%%%%%%%%%%%%%%%%%%%%%%%%%%%%%%%%%%%%%%%%%%%%%%%%%%%%%%%%%%%%%%%%%%%%%%%%%%%%%%%
\section{Introduction}
%%%%%%%%%%%%%%%%%%%%%%%%%%%%%%%%%%%%%%%%%%%%%%%%%%%%%%%%%%%%%%%%%%%%%%%%%%%%%%%%%%%%%%%%%%%%%%%%%%%%%%%%%%%

In 1964 Christenson \etal discovered \CP violation in weak decay~\cite{christensen_1964}. Shortly afterward Sakharov
noted that \CP violation was a crucial ingredient to understanding how our matter dominated universe came into
existence~\cite{sakhaov}.  It was not until 1972 when Kobayashi and Maskawa extended Cabibbo's work on quark mixing to
three generations that \CP violation was introduced into the theory of weak
interactions~\cite{cabibbo,kobayashi_maskawa}. The resulting three generation quark mixing matrix is called the CKM
matrix and this has a single \CP violating phase. Once the magnitude of the elements of this matrix have been measured,
and the \CP violating phase was parameterized by measurement of \CP violation in kaon decays, the CKM matrix could be
used to predict \CP violating effects in other processes. The CKM matrix is
\begin{eqnarray}
\vckm = \theckmmatrix,\nonumber
\end{eqnarray}
and it describes the couplings of the $u$, $c$ and $t$ quarks to $d$, $s$, and $b$ quarks, through transitions mediated
by the exchange of a $W$ boson. In 1981 Bigi and Sanda noted that there could be large \CP violating effects in a
number of \B meson decays, and in particular in the decay of $\Bz \to J/\psi
\ks$~\cite{bigi_sanda_1981}\footnote{Charge conjugation is implied throughout these proceedings.}. At first it was not
obvious how to experimentally test these ideas, that is until Oddone realized that the effects could be observed using
data from collisions at an asymmetric \epem collider~\cite{oddone}. Two asymmetric energy \epem colliders called \B
factories were built to probe \CP violation in \B meson decays, and in doing so, to test the theory behind the CKM
matrix. Recently Kobayashi and Maskawa have been awarded the 2008 Nobel Prize for Physics\footnote{The 2008 prize was
awarded to Nambu, Kobayashi and Maskawa for work on broken symmetries. See {\tt
http://nobelprize.org/nobel\_prizes/physics/laureates/2008/.}} for their contribution to the CKM mechanism.

The remainder of these proceedings describe the accelerators and detectors called \B factories, tests of the CKM theory
through studies of the unitarity triangle via \CP violation and CKM matrix element measurements, tests of \CPT, and
studies of \B mesons decay to final states with two spin-one particles. In these proceedings I summarise one half of
the lectures on experimental results from the \B factories, and the contribution from Bostjan Golob covers the second
half.

%%%%%%%%%%%%%%%%%%%%%%%%%%%%%%%%%%%%%%%%%%%%%%%%%%%%%%%%%%%%%%%%%%%%%%%%%%%%%%%%%%%%%%%%%%%%%%%%%%%%%%%%%%%
\section{The \B factories}
\label{sec:bfactories}
%%%%%%%%%%%%%%%%%%%%%%%%%%%%%%%%%%%%%%%%%%%%%%%%%%%%%%%%%%%%%%%%%%%%%%%%%%%%%%%%%%%%%%%%%%%%%%%%%%%%%%%%%%%

The need to test CKM theory in \B decays led to at least 21 different concepts for \B factories to be
proposed~\cite{dorfan}. Of these only two were built: The \babar experiment~\cite{babarnim} and PEP-II
accelerator~\cite{pepii} at the Stanford Linear Accelerator Center in the USA and the \belle experiment~\cite{bellenim}
and KEKB accelerator~\cite{kekb} at KEK in Japan. The \B factories are similar in design and operation and started to
collect data in 1999, quickly exceeding their original design goals by a large factor. Table~\ref{tbl:datasets} shows
the integrated luminosity recorded at \babar and \belle at various centre of mass energies $\sqrt{s}$. \babar finished
collecting data in 2008 having recorded 433\ifb of data (465\e{6} \BBb pairs), and at the time of writing these
proceedings \belle was still taking data having recorded 1171\ifb of data (1257\e{6} \BBb pairs). These proceedings
discuss experimental measurements made using data taken at the $\Upsilon(4S)$.  The physics process of interest here is
$e^+e^- \to \Upsilon(4S) \to \B \overline{B}$.

\begin{table}[!h]
\begin{center}
\begin{tabular}{|c|ccc|}\hline
               & \babar (\ifb)& Belle (\ifb)& Total (\ifb)\\ \hline
$\Upsilon(5S)$ & \ldots       & 24          & 24 \\
$\Upsilon(4S)$ & 433          & 738         & 1171 \\
$\Upsilon(3S)$ & 30           & \ldots      & 30 \\
$\Upsilon(2S)$ & 14.5         & \ldots      & 14.5 \\
$\Upsilon(1S)$ & \ldots       & 7           & 7 \\
Off-resonance  & 54           & 75          & 129 \\
  \hline
\end{tabular}
\end{center}
\caption{Luminosity of data recorded at different $\sqrt{s}$.}\label{tbl:datasets}
\end{table}

In addition to this {\em interesting} process, there is also a significant cross section for $e^+e^-$ decay into
$q\overline{q}$ where $q$ is a quark lighter than the $b$ quark, and into di-lepton pairs.  These other processes are
backgrounds when studying the decays of $B$ mesons. However, copious amounts of $D$ mesons and $\tau$ leptons are also
created at a $B$ factory: In fact a $B$ factory is really a {\em flavour} factory.

For time-dependent \CP asymmetry measurements, such as those described in Section~\ref{sec:cpviolation}, the $\Bz$ and
$\Bzb$ created in the $\Upsilon(4S)$ decay are in a P wave correlated state. Neutral $B$ mesons can mix\footnote{See
the contribution of U.~Nierst to these proceedings.}, and until one of the $B$ mesons decays, we have only one \Bz and
one \Bzb event in the decay. This EPR correlation stops at the instant one of the $B$ mesons in the event decays. After
that time $t_1$, the other $B$ in the event oscillates between a \Bz and a \Bzb state until it decays at some time
$t_2$.  The difference between these two decay times is used to extract information about \CP violation. In a symmetric
$e^+e^-$ collider time difference corresponds to a spatial separation $\Delta z$ of $30\mu m$ between the $B$ meson
vertices which is too small to be measured in a detector. In an asymmetric energy collider the spatial separation of
vertices is approximately $200\mu m$ which is measurable in a detector.  The need to resolve the two $B$ vertices in an
event is the reason why PEP-II and KEK-B are asymmetric energy $e^+e^-$ colliders.

A $B$ meson that decays into an {\em interesting} final state such as $J/\psi \ks$ is called the \Brec.  The other \B
meson in the event is called the \Btag which is used to determine or {\em tag} the flavour of \Brec at the time that
the first $B$ meson decay occurs.  We don't know which of the \Brec or \Btag decay first, and so the proper time
difference between the decay of the \Brec and \Btag is a signed quantity related to the measured $\Delta z$ by $\Delta
t \simeq \Delta z / c\beta\gamma$.

%%%%%%%%%%%%%%%%%%%%%%%%%%%%%%%%%%%%%%%%%%%%%%%%%%%%%%%%%%%%%%%%%%%%%%%%%%%%%%%%%%%%%%%%%%%%%%%%%%%%%%%%%%%
\section{Unitarity triangle physics}
\label{sec:unitaritytrianglephysics}
%%%%%%%%%%%%%%%%%%%%%%%%%%%%%%%%%%%%%%%%%%%%%%%%%%%%%%%%%%%%%%%%%%%%%%%%%%%%%%%%%%%%%%%%%%%%%%%%%%%%%%%%%%%

The CKM matrix is unitary, so $\vckm \vckm^\dagger = I$, which leads to six complex relations that can each be
represented as closed triangles in the Standard Model (SM). The equation $\vud \vub^* + \vtd \vtb^* + \vcd\vcb^* = 0$
is the one related to the so-called unitarity triangle (shown in Figure~\ref{fig:ut}). This triangle can be completely
parameterised by any two of the three angles $\alpha$, $\beta$, $\gamma$, by measuring the sides, or by constraining
the coordinates of the apex. If we are able to measure more than two of these quantities we can over-constrain the
theory. Sections~\ref{sec:angles:beta} through~\ref{sec:angles:combined} discuss measurements of the angles, and
section~\ref{sec:sidemeasurements} discusses measurements related to the sides of the triangle.  The angles of the
unitarity triangle are given by
\begin{eqnarray}
\alpha \equiv \arg\left[-\vtd\vtb^*/\vud\vub^*\right],  \label{eq:beta}\\
\beta  \equiv \arg\left[ -\vcd\vcb^* / \vtd\vtb^*\right],  \label{eq:alpha} \\
\gamma \equiv \arg\left[ -\vud\vub^* / \vcd\vcb^* \right], \label{eq:gamma}
\end{eqnarray}
and the apex of the unitarity triangle is given by
\begin{equation}
\rhobar + i\etabar \equiv -\frac{\vud\vub^*}{\vcd\vcb^*}.\nonumber
\end{equation}
\begin{figure}[!ht]
\centerline{\includegraphics[width=0.6\textwidth]{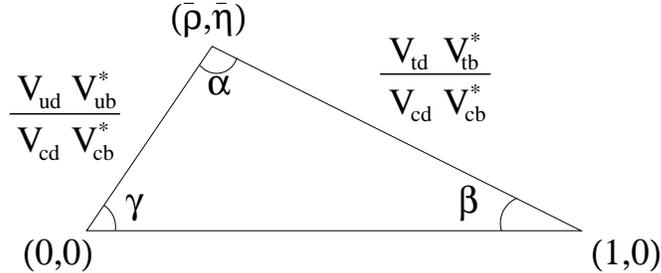}}
\caption{The unitarity triangle.}\label{fig:ut}
\end{figure}
At the current level of experimental precision, we use $\vub = |\vub|e^{i\gamma}$ and $\vtd=|\vtd|e^{i\beta}$.

%%%%%%%%%%%%%%%%%%%%%%%%%%%%%%%%%%%%%%%%%%%%%%%%%%%%%%%%%%%%%%%%%%%%%%%%%%%%%%%%%%%%%%%%%%%%%%%%%%%%%%%%%%%
\subsection{\CP violation measurements}
\label{sec:cpviolation}
%%%%%%%%%%%%%%%%%%%%%%%%%%%%%%%%%%%%%%%%%%%%%%%%%%%%%%%%%%%%%%%%%%%%%%%%%%%%%%%%%%%%%%%%%%%%%%%%%%%%%%%%%%%

The signal decay-rate distribution of a \CP eigenstate decay, $f_+
(f_-)$ for $B_{\rm tag}$= \Bz (\Bzb), is given by:
\begin{eqnarray}
f_{\pm}(\deltat) = \frac{e^{-\left|\deltat\right|/\tau}}{4\tau} [1 \mp \Delta \omega \pm (1 -
2\omega)(-\eta_{f}S\sin(\dmd\deltat) \mp C\cos(\dmd\deltat))] \otimes {\cal R}(\deltat, \sigma(\deltat)),\nonumber
\label{eq:deltatdistribution}
\end{eqnarray}
where $\eta_{f}$ is the \CP eigenvalue of the final state $f$, $\tau=1.530\pm 0.009\ps$ is the mean \Bz\ lifetime and
$\dmd=0.507\pm 0.005 \ps^{-1}$ is the $\Bz-\Bzb$ mixing frequency~\cite{pdg}. The physical decay rate is convoluted
with the detector resolution ${\cal R}(\deltat, \sigma(\deltat))$. As $\Delta \Gamma$ is expected to be small in the
SM, it is assumed that there is no difference between \Bz\ lifetimes, i.e. $\Delta \Gamma = 0$. The parameters $S$ and
$C$ are defined as:
\begin{eqnarray}
 S = \frac{ 2 Im \lambda     }{ 1 + |\lambda|^2}, \hspace{2.0cm}
 C = \frac{ 1 - |\lambda|^2 }{ 1 + |\lambda|^2},\nonumber
\end{eqnarray}
where $\lambda=\frac{q}{p}\frac{\overline{A}}{A}$ is related to the level of \Bz-\Bzb\ mixing ($q/p$), and the ratio of
amplitudes of the decay of a \Bzb\ or \Bz\ to the final state under study ($\overline{A}/A$). Sometimes we assign the
wrong flavour to \Btag.  The probability for this to happen is given by the mis-tag fraction $\omega$, where $\Delta
\omega$ is the difference between the mistag probability of \Bz and \Bzb decays.

\CP violation is probed by studying the time-dependent decay-rate asymmetry
\begin{equation}
 {\cal A} = \frac{
  R (\deltat) - \overline{R}(\deltat) } { R (\deltat) + \overline{R}(\deltat) } = -\eta_{f}S\sin(\dmd\deltat)-C\cos(\dmd\deltat),\nonumber
\end{equation}
where $R$($\overline{R}$) is the decay-rate for \Bz(\Bzb) tagged events. The Belle Collaboration use a different
convention to that of the \babar\ Collaboration with $C=-A_{\CP}$. Here all results are quoted using the $S$ and $C$
convention.

In the case of charged \B-meson decays (and $\pi^0\pi^0$ as there is no vertex information) one can study a time
integrated charge asymmetry
\begin{equation}
A_{\CP}=\frac{\overline{N}-N}{\overline{N}+N},\nonumber
\end{equation}
where $N$ ($\overline{N}$) is the number of \B\ ($\overline {\B}$) decays to the final state. A non-zero measurement of
$S$, $C$ or $A_{\CP}$ is a clear indication of \CP violation.

In order to quantify the mistag probabilities and resolution function parameters, the \B factories study decay modes to
flavour specific final states.  These states form what is usually referred to as the \Bflav sample of events.  The
following decay modes are included in the \Bflav sample: $\B \to D^{(*)-}\pi^+$, $D^{(*)-}\rho^+$, and $D^{(*)-}a_1^+$.
It is assumed that the mistag probabilities and resolution function parameters determined for the \Bflav sample are the
same as those for the signal \Brec decays.

There are three types of \CP Violation that can occur: (i) \CP Violation in mixing, which requires $|q/p| \neq 1$, (ii)
CP Violation in decay (also called direct \CP violation) where $|\overline{A}/A| \neq 1$, and (iii) \CP violation in
the interference between mixing and decay amplitudes.

%%%%%%%%%%%%%%%%%%%%%%%%%%%%%%%%%%%%%%%%%%%%%%%%%%%%%%%%%%%%%%%%%%%%%%%%%%%%%%%%%%%%%%%%%%%%%%%%%%%%%%%%%%%
\subsubsection{The angle $\beta$}
\label{sec:angles:beta}
%%%%%%%%%%%%%%%%%%%%%%%%%%%%%%%%%%%%%%%%%%%%%%%%%%%%%%%%%%%%%%%%%%%%%%%%%%%%%%%%%%%%%%%%%%%%%%%%%%%%%%%%%%%

The golden channel predicted to be the best one to observe \CP violation in \B meson decays through the measurement of
\sintwobeta is $\Bz \to J/\psi \ks$~\cite{bigi_sanda_1981}. The phase $\beta$ comes from the \vtb vertices of the
$\Bz-\overline{B}^0$ mixing amplitudes. This is just one of the theoretically clean $b\to \ccbars$ Charmonium decays,
where the measurement of $S$ is a direct measurement of \sintwobeta, neglecting the small effect of mixing in the
neutral kaon system. The other theoretically clean decays include $\psi(2S)\ks$, $\chi_{1c}\ks$, $\eta_c\ks$, and
$J/\psi K^{*0}$. Figure~\ref{fig:ccsfeynman_diagrams} shows the mixing and tree diagrams relevant for $b\to \ccbars$
Charmonium decays.
\begin{figure}[!h]
\begin{center}
\resizebox{12.5cm}{!}{
  \includegraphics{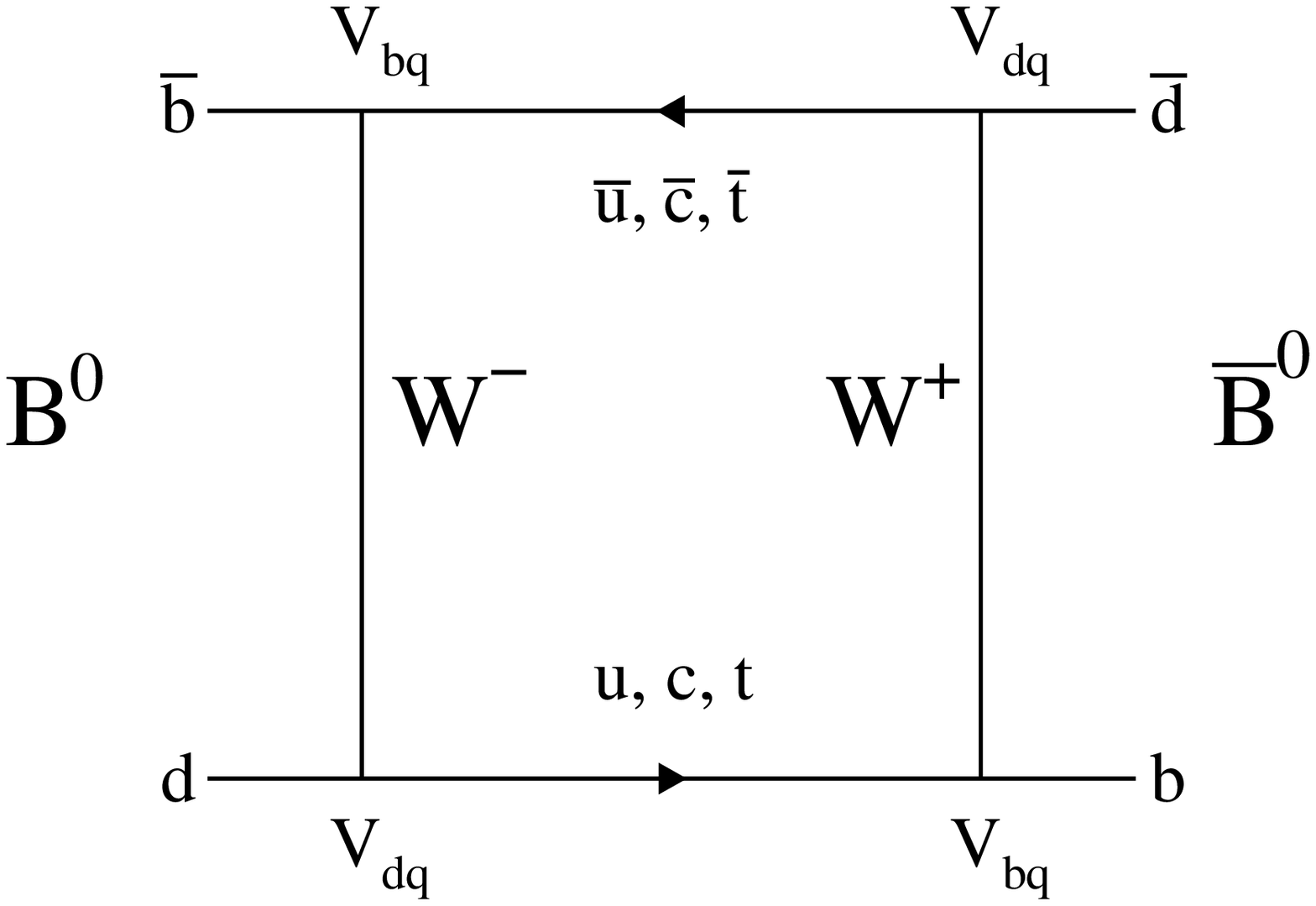}
  \includegraphics{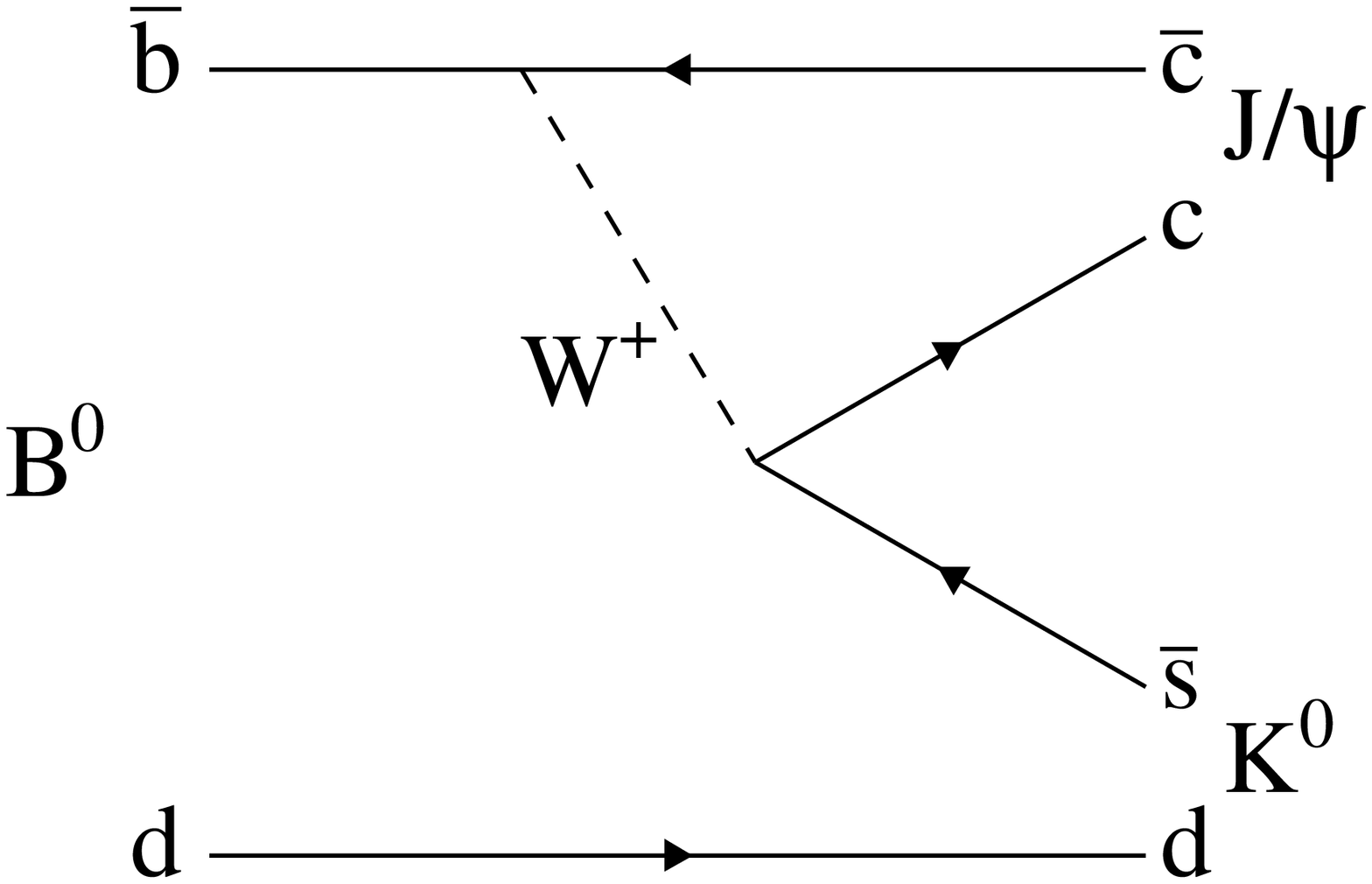}
  }
\end{center}
 \caption{The (left) mixing and (right) tree contributions to Charmonium decays.}
\label{fig:ccsfeynman_diagrams}
\end{figure}
There are several calculations of the level of SM uncertainty on the measurement of \sintwobeta in $b\to \ccbars$
decays which include theoretical and data driven phenomenological estimates of this
uncertainty~\cite{sin2beta:theory_err:li,sin2beta:theory_err:boos,sin2beta:theory_err:ciuchini}. The data driven method
uses $\Bz \to J/\psi \pi^0$ to limit the SM uncertainties at a level of $10^{-2}$, and the theoretical calculations
limit these uncertainties to be ${\cal O}(10^{-3})$ to ${\cal O}(10^{-4})$. \babar found a signal for \CP violation in
\B meson decay in 2001~\cite{ref:sin2beta:babardiscovery} and this result was confirmed two weeks later by
\belle~\cite{ref:sin2beta:belleconfirmation}. The latest analyses from the \B factories provide the most precise test
of CKM theory~\cite{ref:sin2beta:babar,ref:sin2beta:belle}. These results are summarised in
Table~\ref{tbl:sintwobetaresults} where the \babar result uses $B$ decays to $J/\psi K^0$, $\psi(2S) K^0_S$,
$\chi_{1c}K^0_S$, $\eta_{c} K^0_S$, and $J/\psi  K^{*0}$ to measure \sintwobeta.  \belle use $J/\psi K^0$, and
$\psi(2S) K^0_S$ final states for their measurement.

\begin{wraptable}{r}{0.6\textwidth}
\centerline{\begin{tabular}{|l|c|} \hline
Experiment & \sintwobeta \\ \hline
\babar & $0.691 \pm 0.029 \stat \pm 0.014 \syst$\\
\belle & $0.650 \pm 0.029 \stat \pm 0.018 \syst$\\
World Average & $0.671 \pm 0.024$ \\ \hline
\end{tabular}}
\caption{Experimental results for \sintwobeta from the \B factories.}
\label{tbl:sintwobetaresults}
\end{wraptable}
When converting the measured value of \sintwobeta to a value of $\beta$ we obtain a four fold ambiguity on $\beta$. The
four solutions for beta are $21.1^\circ$, $68.9^\circ$, $201.1^\circ$, and $248.9^\circ$.  The two solutions
$68.9^\circ$ and $248.9^\circ$ are disfavoured by $\cos 2\beta$ measurements from decays such as $\Bz \to J/\psi
K^{*}$~\cite{cos2beta:jpsikstar}, $D^{*}D^{*}K_S^0$~\cite{cos2beta:dstdstks} and $D^{*0} h^0$~\cite{babar:dzeroh}. The
only solution for $\beta$ that is consistent with the Standard Model is $\beta = (21.1\pm 0.9)^\circ$. This result
corresponds to the first test of the CKM mechanism as the apex of the unitarity triangle can be constrained using
Eq.~\ref{eq:beta}. In order to fully constrain the theory, we need a second measurement from one of the observables
described below.

As can be seen from Table~\ref{tbl:sintwobetaresults}, the precision of the \sintwobeta result from the \B factories is
still limited by statistics. This measurement will be refined by the next generation of experiments, including
LHCb~\cite{lhcb}, SuperB~\cite{superb} and SuperKEKB~\cite{superkekb}. For example, the measurement of \sintwobeta with
75\iab from SuperB will be systematics limited and have a precision of $\pm 0.005$~\cite{superb}.

%%%%%%%%%%%%%%%%%%%%%%%%%%%%%%%%%%%%%%%%%%%%%%%%%%%%%%%%%%%%%%%%%%%%%%%%%%%%%%%%%%%%%%%%%%%%%%%%%%%%%%%%%%%
\subsubsection{The angle $\alpha$}
\label{sec:angles:alpha}
%%%%%%%%%%%%%%%%%%%%%%%%%%%%%%%%%%%%%%%%%%%%%%%%%%%%%%%%%%%%%%%%%%%%%%%%%%%%%%%%%%%%%%%%%%%%%%%%%%%%%%%%%%%
The measurement of $\alpha$ is not as straight forward as $\beta$. All of the decay channels that are sensitive to
$\alpha$ have potentially large contributions from loop amplitudes\footnote{These loop amplitudes are often called
'penguins' in \B physics literature.  This nomenclature stems from a lost bet as described in Ref.~\cite{penguins}.},
in addition to the leading order tree and mixing contributions.  Figure~\ref{fig:hhfeynman_diagrams} shows these tree
and loop contributions.  In the absence of a loop contribution, the interference between tree and mixing amplitudes
would result in $S=\sin(2\alpha)$. Here the weak phase\footnote{A weak phase is one that changes sign under \CP. The
angles of the unitarity triangle are weak phases.} measured is $\alpha = \pi - \beta - \gamma$ where $\beta$ comes from
the \vtd vertices of the mixing amplitudes, and $\gamma$ comes from the \vub vertex of the tree amplitude. However, the
loop contributions have a different weak phase to the tree contribution, so they 'pollute' the measurement of $\alpha$.
There are two schemes used in order to determine the loop pollution $\delta \alpha$: (i) use \su{2}
relations~\cite{gronaulondon}, and (ii) use \su{3} relations~\cite{beneke_su3} to constrain the effect of loop
amplitudes on the extraction of $\alpha$ from measurements of \B meson decays to $h^+h^-$ final states, where $h=\pi,
\rho$. The effect of loop amplitudes is $\delta \alpha = \alpha - \alphaeff$, where \alphaeff is related to the
measured $S$ and $C$ via $S = \sqrt{1 - C^2}\sin(2 \alphaeff)$.

\begin{figure}[!h]
\begin{center}
  \resizebox{10.0cm}{!}{\includegraphics{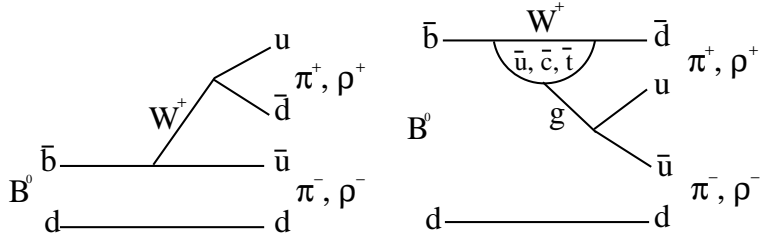}}
\end{center}
 \caption{The tree (left) and gluonic loop (right) contributions to $B \to h^+h^-$ decays.}
\label{fig:hhfeynman_diagrams}
\end{figure}

One can use \su{2} isospin to relate the amplitudes of \B decays
to $\pi\pi$ final states~\cite{gronaulondon}.  This results in two
relations:
\begin{eqnarray}
\frac{1}{\sqrt{2}}A^{+-}=A^{+0}-A^{00}, \,\,\,\,\,\,\,
\frac{1}{\sqrt{2}}\overline{A}^{+-}=\overline{A}^{-0}-\overline{A}^{00},\nonumber
\end{eqnarray}
where $A^{ij}$ ($\overline{A}^{ij}$) are the amplitudes of \B\ ($\overline{B}$) decays to the final state with charge
$ij$. These two relations correspond to triangles in a complex plane with a common base given by $|A^{+0}| =
|\overline{A}^{-0}|$ neglecting electroweak loop contributions. There are three such relations for $\rho\rho$ decays,
one for each of the transversity states (Section~\ref{sec:btovv}). The extraction of $\alpha$ from $\rho\pi$ decays is
complicated by the fact that the final state is not a \CP eigenstate~\cite{lipkin,snyderquinn}.  A more detailed
overview of the experimental methods used for $\pi\pi$, $\rho\rho$, and $\rho\pi$ decays is given in
Ref.~\cite{alphareview}.

The experimental results for $\B \to \pi\pi$~\cite{ref:alphapipi:belle,ref:alphapipi:babar}, $\B \to
\rho\rho$~\cite{ref:alpharhoprhoz:babar,ref:alpharhoprhom:babar,ref:alpharhozrhoz:babar,ref:alpharhoprhoz:belle,ref:alpharhoprhom:belle,ref:alpharhozrhoz:belle},
and $\B \to \rho\pi$~\cite{ref:alpharhopi:babar,ref:alpharhopi:belle} decays can be combined together in order to
constrain $\alpha$.  This constraint is shown in Figure~\ref{fig:angles:alpha} for the \su{2} approach. The solution
compatible with the SM is $\alpha = (91\pm 8)^\circ$~\cite{utfit} which provides a second reference point to test the
CKM mechanism.

Beneke \etal proposed the use of \su{3} to relate the loop component of $B^+ \to \rho^- K^{*0}$ to the loop component
of $\Bz\to \rho^+\rho^-$~\cite{beneke_su3}.  In order to do this, one has to measure the branching fractions and
fractions of longitudinally polarized signal in both decay channels, as well as $S$ and $C$ for the longitudinal
polarization of $\Bz\to \rho^+\rho^-$.  On doing this, \babar finds that $\alpha=(89.8^{+7.0}_{-6.4})^\circ$, where the
corresponding loop to tree ratio measured is $0.10^{+0.03}_{-0.04}$~\cite{ref:alpharhoprhom:babar}.

\begin{wrapfigure}{r}{0.5\textwidth}
\centerline{\includegraphics[width=0.5\textwidth]{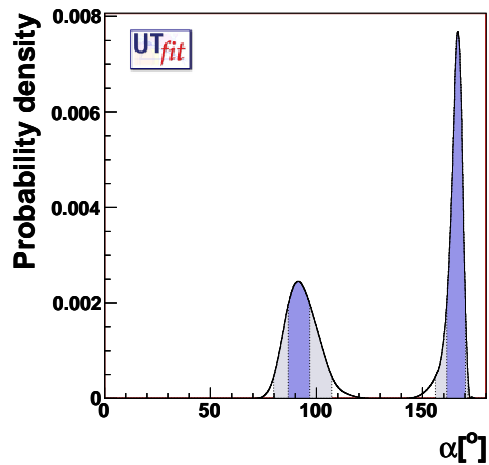}} \caption{The constraint on $\alpha$ from an
isospin analysis of $\B\to hh$ decays.  Constraints are made on the magnitude of the penguin to tree ratio for $B\to
\pi\pi$ decays when making this plot. This figure is reproduced from UT fit~\cite{utfit}. The shaded regions correspond
to the allowed
 solutions for $\alpha$.}\label{fig:angles:alpha}
\end{wrapfigure}

The strongest constraint on $\alpha$ comes from the study of $\B \to \rho\rho$ decays and this measurement is currently
limited by statistics.  The next generation of experiments will be able to refine our knowledge of $\alpha$: LHCb will
be able to measure this to ${\cal O}(5^\circ)$~\cite{lhcbalpha} with 10\ifb of data using $\B \to \rho\pi$ decays, but
will not be able to measure all of the necessary inputs for the $\pi\pi$ and $\rho\rho$ measurements. The SuperB
experiment will be able to measure $\alpha$ to a level that will be limited by systematic and theoretical
uncertainties: ${\cal O}(1-2^\circ)$~\cite{superb} with a data sample of 75\iab.

It is also possible to constrain $\alpha$ using \su{3} based approaches for decays such as $B\to a_1 \pi$, and
$a_1\rho$. Even though these decays are experimentally challenging to measure, the time-dependent analysis of $B\to a_1
\pi$ decays has been performed~\cite{babar:aonepi}.  Additional experimental constraints, such as the branching
fractions of $K_1\pi$ decays, are required to interpret those results as a measurement on $\alpha$.  Only a branching
fraction upper limit exists for $B^0\to a_1^\pm \rho^\mp$~\cite{babar:aonerho}.

%%%%%%%%%%%%%%%%%%%%%%%%%%%%%%%%%%%%%%%%%%%%%%%%%%%%%%%%%%%%%%%%%%%%%%%%%%%%%%%%%%%%%%%%%%%%%%%%%%%%%%%%%%%
\subsubsection{The angle $\gamma$}
\label{sec:angles:gamma}
%%%%%%%%%%%%%%%%%%%%%%%%%%%%%%%%%%%%%%%%%%%%%%%%%%%%%%%%%%%%%%%%%%%%%%%%%%%%%%%%%%%%%%%%%%%%%%%%%%%%%%%%%%%

There are several promising methods being pursued in order to constrain $\gamma$ or $\sin(2\beta + \gamma)$, however
none of these provides as stringent a bound as those for $\beta$ and $\alpha$.  Here I discuss three methods used to
constrain $\gamma$: these are called Gronau-London-Wyler (GLW)~\cite{ref:glw}, Attwood-Dunietz-Soni
(ADS)~\cite{ref:ads} and Giri-Grossman-Soffer-Zupan (GGSZ)~\cite{ref:ggsz}. These three methods are theoretically
clean, and use $B$ decays to $D^{(*)}K^{(*)}$ final states to measure $\gamma$.

The GLW method~\cite{ref:glw} uses $B^+\to D_{\CP}^0X^+$ and $B^+\to \overline{D}_{\CP}^0X^+$ where $X^+$ is a
strangeness one state, and $D_{\CP}^0$ is a $D^0$ decay to a \CP eigenstate (similarly for $\overline{D}_{\CP}^0$) to
extract $\gamma$.  The \CP-even eigenstates used are $D_{\CP}^0 \to h^+h^-$ where $h=\pi, K$, and the \CP-odd
eigenstates used are $D_{\CP}^0 \to \ks \pi^0, \ks\omega$, and $\ks \phi$. The ratio of Cabibbo allowed to Cabibbo
suppressed decays is given by the parameter $r_B$.  The experimentally determined value is $r_B\sim 0.1$ which leads to
a relatively a large uncertainty on $\gamma$ extracted using this method. A similar measurement has been performed
using $D^{*0}_{\CP}K$ decays, where $r_B^*$ is found to be $0.22\pm 0.09\pm 0.03$, and only weak constraints can be
placed on $\gamma$~\cite{babar:glw,belle:glw}.

The ADS method~\cite{ref:ads} uses the doubly Cabibbo suppressed decays $B^+ \to D^{*0} K^{(*)\pm}$ where the
interference between amplitudes with a $D$ and a $\overline{D}$ decaying into a $K^+\pi^-$ final state is sensitive to
$\gamma$. As with the GLW method, the ADS method requires more statistics than are currently available in order to
measure $\gamma$~\cite{belle:ads,babar:ads}.

The GGSZ method~\cite{ref:ggsz} uses $B$ decays to $D^{(*)0} K^{(*)}$ final states where the $D^{(*)}$ subsequently
decays to $\ks h^+h^-$ ($h = \pi, K$)to constrain $\gamma$.  This method is self tagging either by the charge of the
reconstructed $B^\pm$ meson, or by the charge of the reconstructed $K^{(*)}$ for neutral $B$ decays. One has to
understand the $D$ Dalitz decay distribution to determine $\gamma$. Using this method \belle measure $\gamma =
(76^{+12}_{-13} \pm 4 \pm 9)^\circ$~\cite{belle:ggsz} where the errors are statistical, systematic and model dependent.
The corresponding \babar measurement is $\gamma = (76^{+23}_{-24} \pm 5 \pm 5)^\circ$~\cite{babar:ggsz}. The difference
in statistical uncertainties of these measurements comes from the fact that \belle measure a larger value of $r_B$ than
\babar.

Figure~\ref{fig:angles:gamma} shows the experimental constraint on $\gamma$ where the total precision on this angle is
$20^\circ$ with a central value of $71^\circ$.  The next generation of experiments will be able to refine our knowledge of $\gamma$: LHCb will be able to
measure this to ${\cal O}(2^\circ)$~\cite{lhcbalpha} with 10\ifb of data. The SuperB experiment will be able to measure
$\gamma$ to ${\cal O}(1^\circ)$~\cite{superb} with a data sample of 75\iab.

\begin{figure}[!h]
\begin{center}
\resizebox{10.0cm}{!}{
  \includegraphics{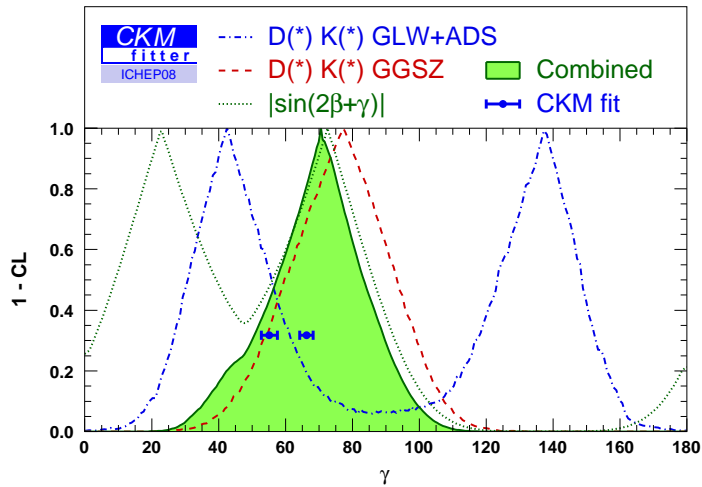}
  }
\end{center} \caption{The experimental constraint on $\gamma$.  This figure is from CKM
Fitter~\cite{ckmfitter}.}\label{fig:angles:gamma}
\end{figure}

\subsubsection{Angle constraints on the CKM theory}
\label{sec:angles:combined}

The angle measurements described in the previous sections individually constrain CKM theory by restricting the allowed
values of \rhobar and \etabar.  The individual and combined constraints of these measurements are shown in
Fig.~\ref{fig:utconstraints:angles}.  The angle constraints are consistent with CKM theory at the current level of
precision.  Combining the angle measurements: $\beta=(21.1\pm 0.9)^\circ$, $\alpha=(89.9^{+7.0}_{-6.4})^\circ$, and
$\gamma=(71\pm 20)^\circ$, we obtain $\rhobar = 0.13 \pm 0.04$ and $\etabar = 0.34 \pm 0.02$.  The the precision on
these constraints is dominated by our knowledge of $\alpha$ and $\beta$. CKM theory requires that $\alpha+\beta+\gamma
= 180^\circ$. The \B factory measurements give $\alpha+\beta+\gamma = (190 \pm 21)^\circ$ where the precision of this
test is limited by our knowledge of $\gamma$.

\begin{figure}[!h]
\begin{center}
  \resizebox{14.5cm}{!}{
  \includegraphics{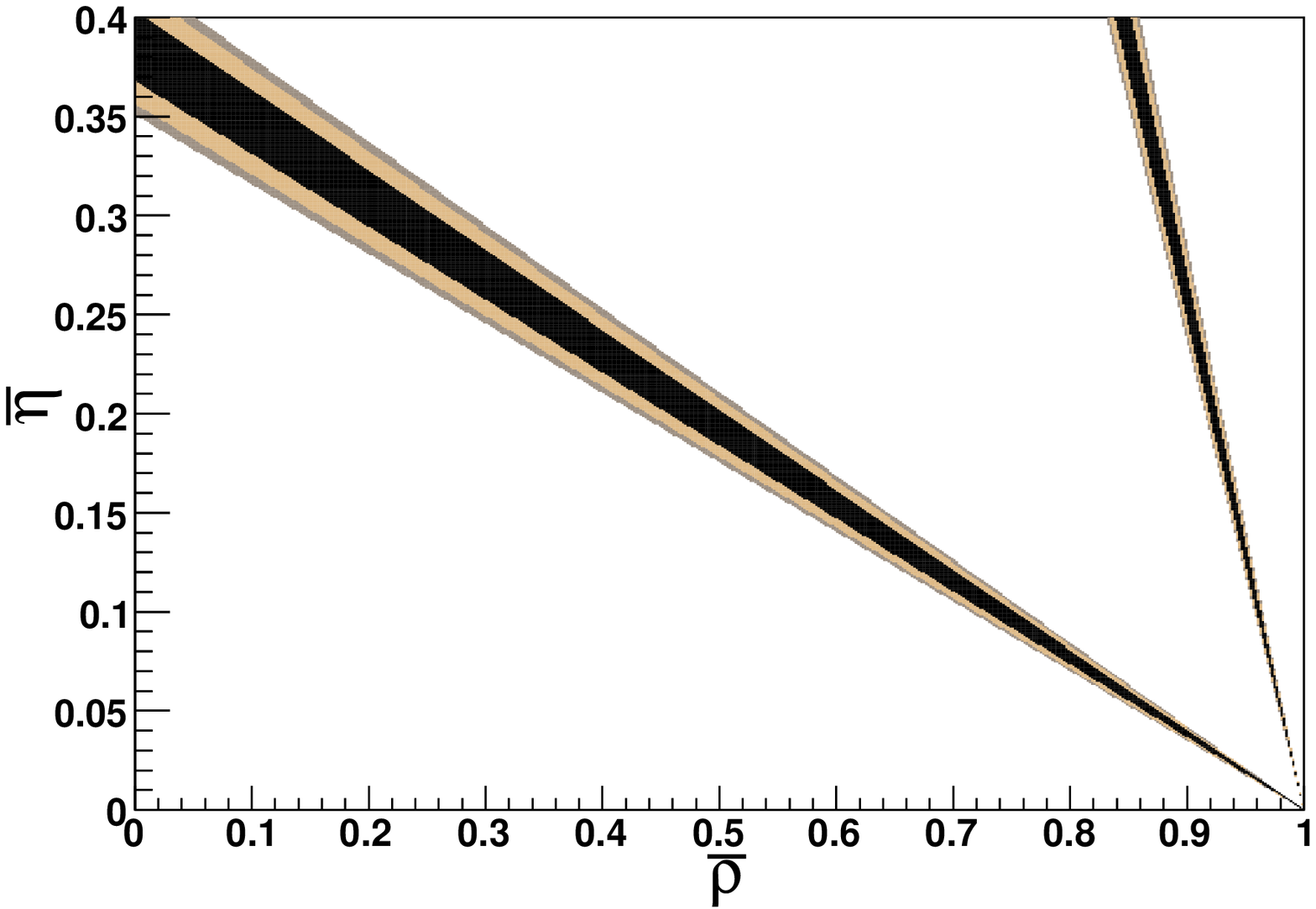}
  \includegraphics{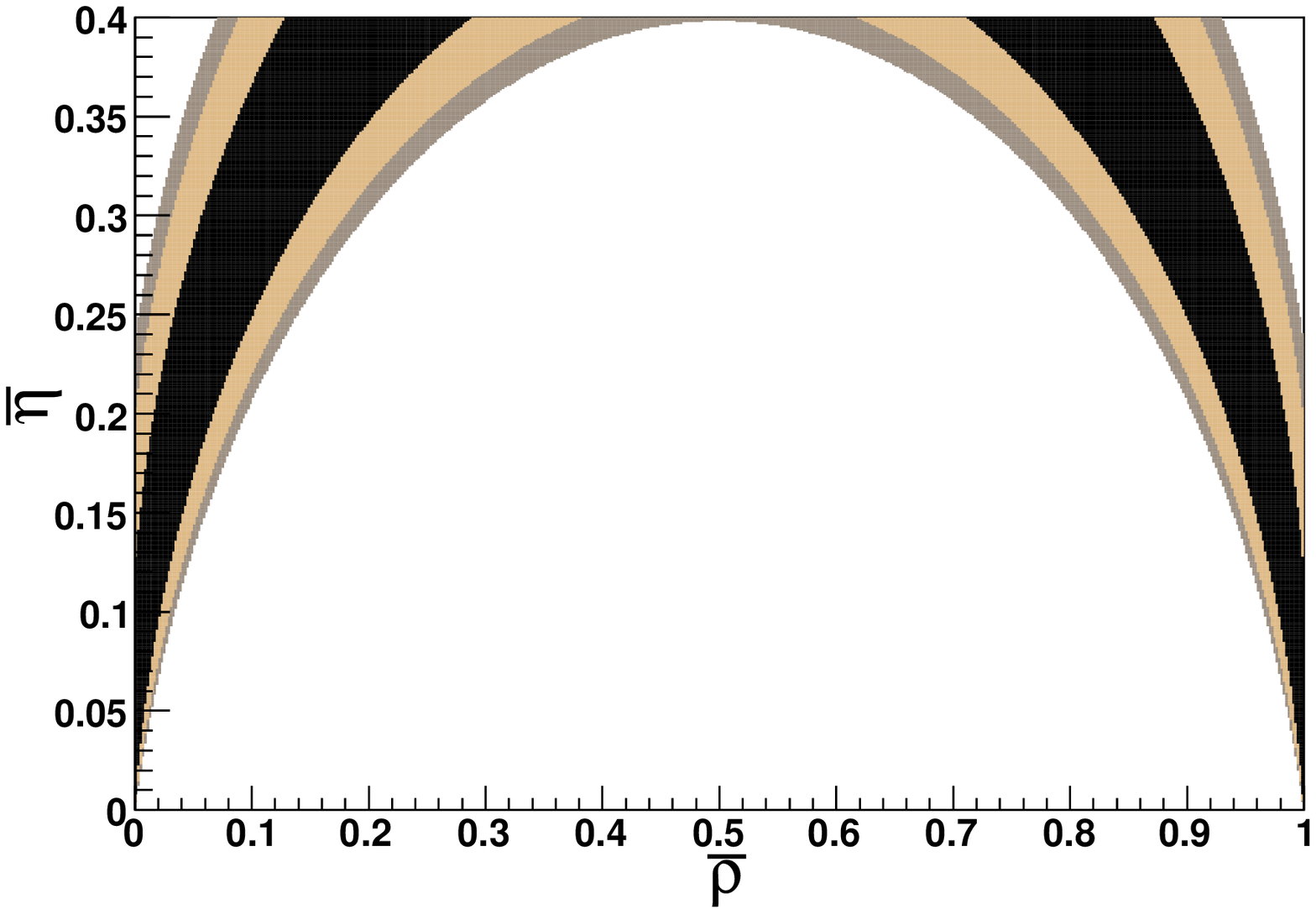}
}
  \resizebox{14.5cm}{!}{
  \includegraphics{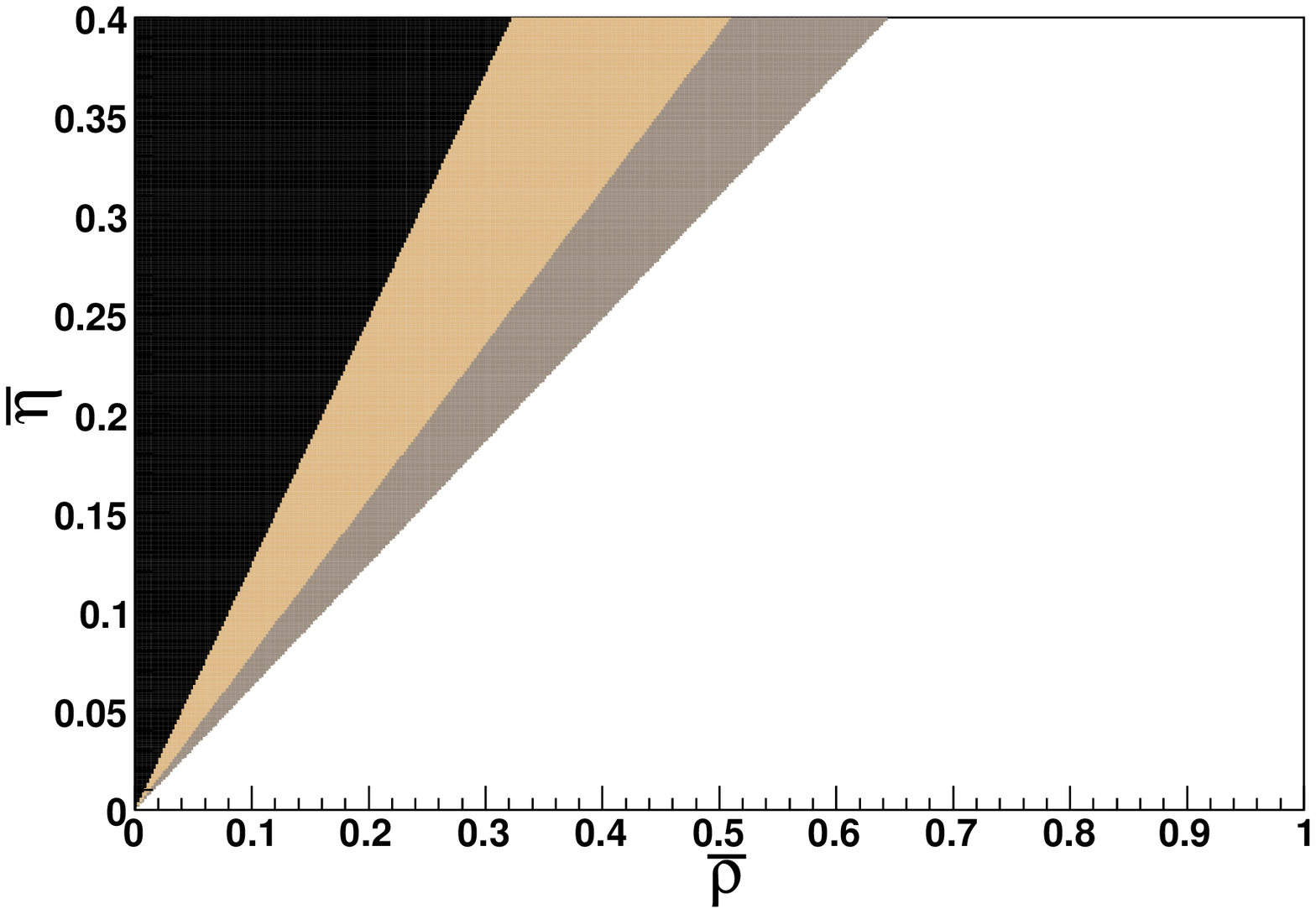}
  \includegraphics{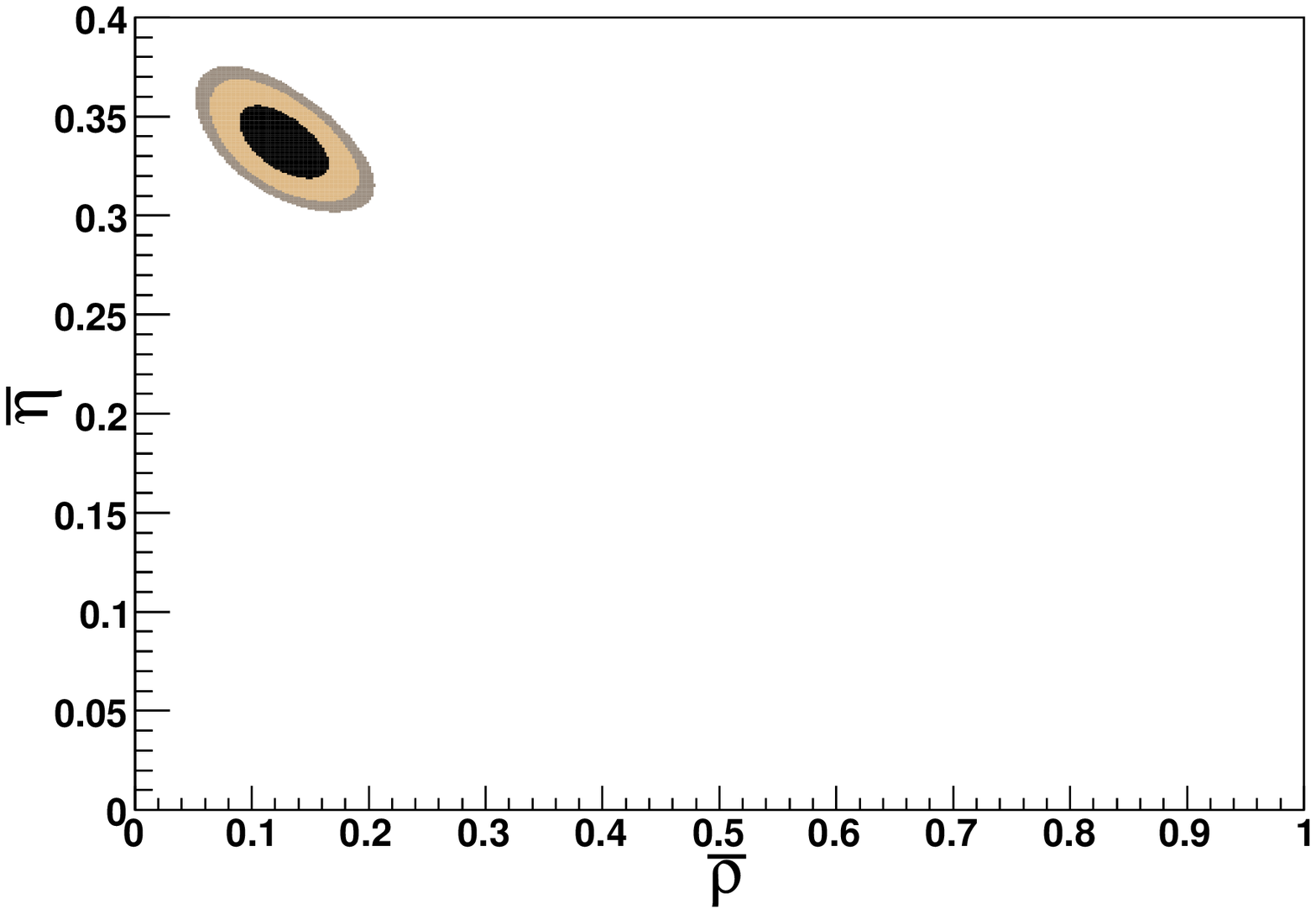}
  }
\end{center}
 \caption{Angle constraints on \rhobar and \etabar from (top left) $\beta$, (top right) $\alpha$,
 (bottom left) $\gamma$, and (bottom right) combined.  The shaded contours show the 68\% (black), 90\% (light) and 95\% (medium) confidence levels.} \label{fig:utconstraints:angles}
\end{figure}

%%%%%%%%%%%%%%%%%%%%%%%%%%%%%%%%%%%%%%%%%%%%%%%%%%%%%%%%%%%%%%%%%%%%%%%%%%%%%%%%%%%%%%%%%%%%%%%%%%%%%%%%%%%
\subsubsection{Direct \CP violation}
%%%%%%%%%%%%%%%%%%%%%%%%%%%%%%%%%%%%%%%%%%%%%%%%%%%%%%%%%%%%%%%%%%%%%%%%%%%%%%%%%%%%%%%%%%%%%%%%%%%%%%%%%%%

\begin{wraptable}{r}{0.6\textwidth}
\centerline{\begin{tabular}{|l|c|} \hline
 Experiment & $A_{\CP}$ \\ \hline\noalign{\vspace{1pt}}
 \babar & $-0.107 \pm 0.016^{+0.006}_{-0.004}$~\cite{babar:latestkpi}\\
 \belle & $-0.094\pm 0.018\pm 0.008$~\cite{belle:kpi}\\
 CDF    & $-0.086\pm 0.023\pm 0.009$~\cite{cdf:kpi}\\
 CLEO   & $-0.04\pm 0.16\pm 0.02$~\cite{cleo:kpi}\\\hline
\end{tabular}}
 \caption{Experimental results for $A_{\CP}$, where the first error quoted is statistical and the second is
 systematic.}
 \label{tbl:kpi}
\end{wraptable}
Direct \CP violation was established by the NA48 and KTeV experiments in 1999~\cite{na48,ktev} through the measurement
of a non-zero value of the parameter $\epsilon^\prime/\epsilon$.  This phenomenon was confirmed 45 years after \CP
violation was discovered in kaon decays.  In contrast to this in \B meson decay direct \CP violation was observed
only a few years after \CP violation was established.  The first observation of direct \CP violation in \B decays was
made via the measurement of a non-zero $A_{\CP}$ in $B^0 \to K^\pm \pi^\mp$ decays in 2007 by \babar~\cite{babar:kpi}.
The following year \belle confirmed this result~\cite{belle:kpi}.  The latest results of this measurement are
summarised in Table~\ref{tbl:kpi}.  It has been suggested that the difference in the direct \CP violation observed in
$\Bz \to K^\pm \pi^\mp$ and $\B^+ \to K^+\pi^0$ could be due to new physics (See Ref.~\cite{babar:kpi} and references
therein).  A more plausible explanation is that the difference arises from final state interactions~\cite{ref:chuakpi}.

%%%%%%%%%%%%%%%%%%%%%%%%%%%%%%%%%%%%%%%%%%%%%%%%%%%%%%%%%%%%%%%%%%%%%%%%%%%%%%%%%%%%%%%%%%%%%%%%%%%%%%%%%%%
\subsubsection{Searching for new physics}
%%%%%%%%%%%%%%%%%%%%%%%%%%%%%%%%%%%%%%%%%%%%%%%%%%%%%%%%%%%%%%%%%%%%%%%%%%%%%%%%%%%%%%%%%%%%%%%%%%%%%%%%%%%

The \B factories have seen evidence for, or have observed indirect \CP violation in $\Bz\to J/\psi \kz$, $\Bz\to J/\psi
\pi^0$, $\Bz\to \psi(2S) \ks$, $\Bz\to \eta_{1c} \ks$, $\Bz\to \eta^\prime \kz$, $\Bz\to f_0^0(980) \ks$, $\Bz \to
K^+K^-\kz$, $\Bz \to D^{*+}D^{*-}$, and $\Bz\to \pi^+\pi^-$.  They have also seen evidence for or observed direct \CP
violation in $\Bz\to \pi^+\pi^-$, $\Bz\to \eta K^{*0}$, $\Bz\to \rho^\pm\pi^\mp$, $\Bz\to K^\pm \pi^\mp$, $\Bpm \to
\rho^0 K^\mp$, $B\to D^0_{\CP+}K$, $B\to D^{(*)0}K^*$.  All of the measurements of \CP violating asymmetries to date
are consistent with CKM theory.  It is possible that there is more to \CP violation than the CKM theory and the rest of
this section discusses one way to search for effects beyond CKM.

\begin{wrapfigure}{r}{0.6\textwidth}
\centerline{\includegraphics[width=0.6\textwidth]{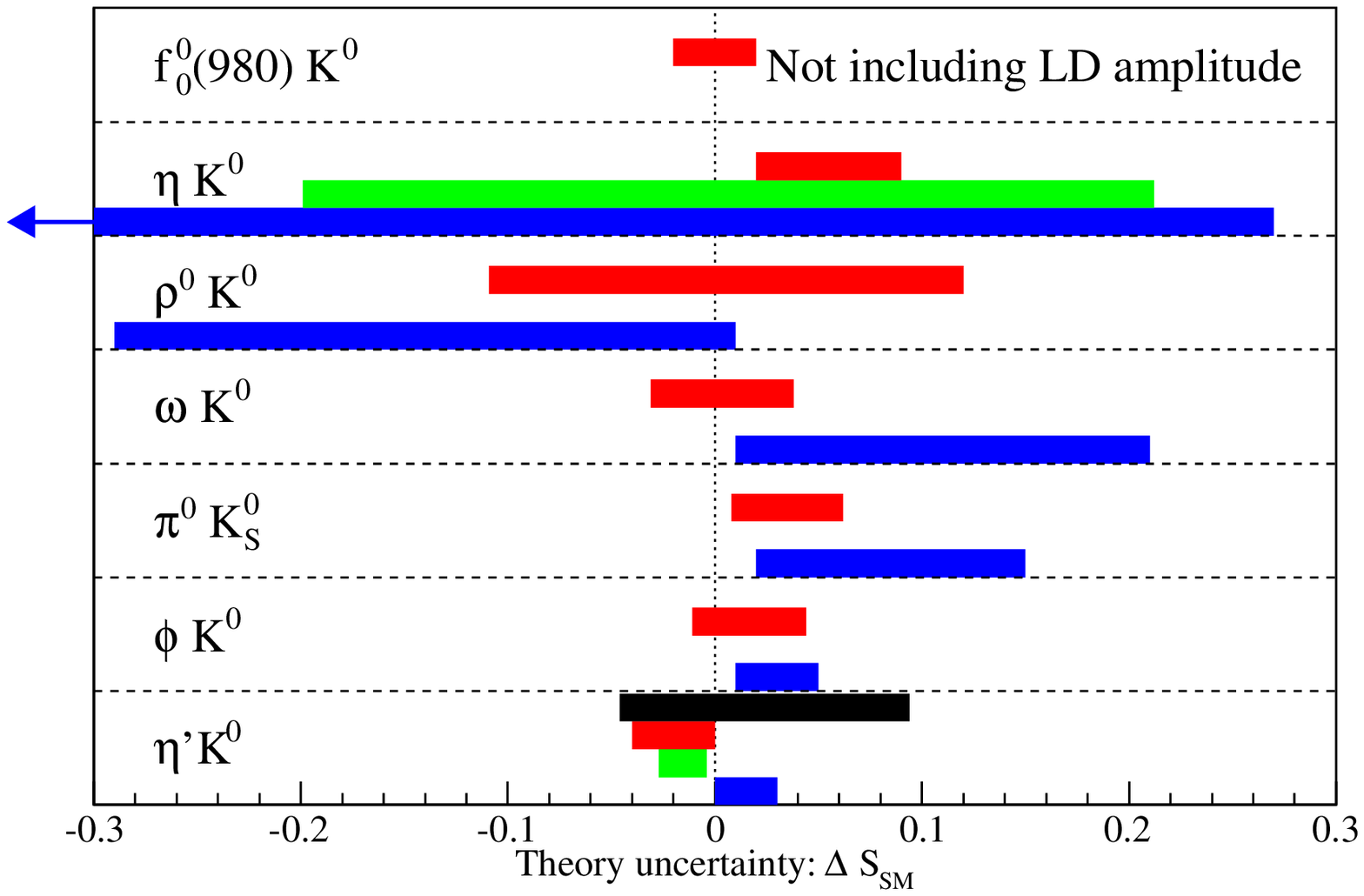}} \caption{Theoretical estimates of
\deltassm.}\label{fig:deltastheory}
\end{wrapfigure}
A large number of rare \B decays are sensitive to $\beta$ however as these measurements are not necessarily clean we
call the phase measured $\betaeff$. These fall into two categories: those that are loop dominated; and those that have
a loop and a tree contribution. The SM loop amplitude can be replaced by a corresponding amplitude with unknown heavy
particles, for example the SUSY partners of the SM loop constituents, so the loops are sensitive to the presence of new
physics.  The consequence of this is that if there are new heavy particles that contribute to the loop, the SM
calculated expectation for observables will differ from experimental measurements of the observables. \sintwobeta has
been measured to an accuracy of $1^\circ$ using tree dominated \ccbars decays and this can be used as a reference point
to test for deviations from the SM.  If we measure $\sintwobetaeff$ for a rare decay, then $\deltas =
\sintwobetaeff-\sintwobeta -\deltassm$ is zero in the absence of new physics.  Here $\deltassm$ is a term that accounts
for the effect of possible higher order SM contributions to a process that would lead to the measured $\sintwobetaeff$
differing from the \ccbars measurement. Such effects include contributions from long distance scattering (LD),
annihilation topologies and other often neglected terms. There has been considerable theoretical effort in recent years
to try and constrain \deltassm which is summarised in Figure~\ref{fig:deltastheory}. The figure is divided into decay
modes, and each decay mode has up to four error bands drawn on it.  These error bands come from (top to bottom)
calculations by Beneke \etal~\cite{spenguin:theory:beneke}, Williamson and
Zupan~\cite{spenguin:theory:williamson_zupan}, Cheng \etal~\cite{spenguin:theory:cheng}, and Gronau
\etal~\cite{spenguin:theory:gronau}.  It is clear from this work that some of the decay modes are clean, and the SM
expectation of \sintwobetaeff is essentially the same as the expectation for \sintwobeta from \ccbars.  However some
modes have a significant contribution to \deltas from \deltassm. The experimental situation is shown in
Figure~\ref{fig:deltasexpt}~\cite{alternatebeta:expt}.

\begin{wrapfigure}{r}{0.6\textwidth}
\centerline{\includegraphics[width=0.6\textwidth]{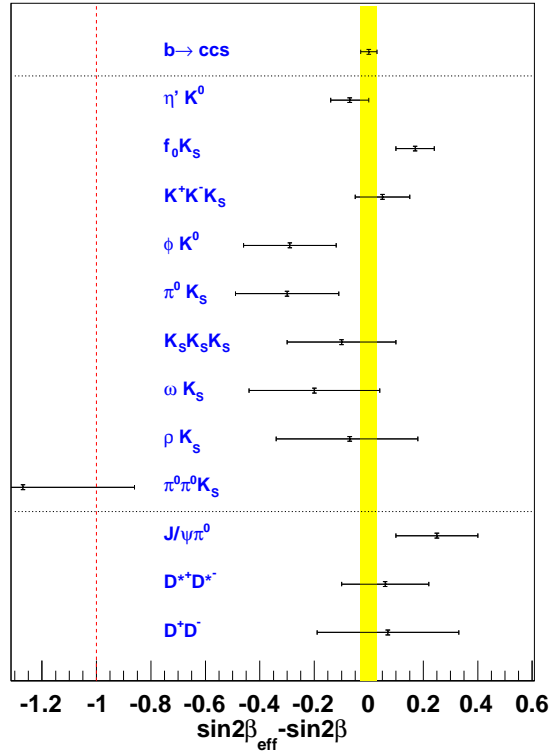}} \caption{Measurements of
\sintwobetaeff-\sintwobeta.  The top part and vertical band show the reference measurement from \ccbars decays (See
Sec.~\ref{sec:angles:beta}), the middle part shows measurements from $b \to s$ loop processes, and the part section
shows measurements from $b\to d$ processes with loop and tree contributions. All results shown are averages of
measurements from \babar and \belle.}\label{fig:deltasexpt}
\end{wrapfigure}

The most precisely determined \sintwobetaeff from a $b\to s$ loop process is that of $\Bz \to \etap \kz$. This is also
one of the theoretically cleanest channels, and is consistent with the SM expectation of $\deltas=0$ at the current
precision.  In recent years it has been frequently noted that the average value of $\sintwobetaeff-\sintwobeta$ is less
than zero with a significance of between two and three standard deviations. However as discussed above, it is not
correct to compare the average of any set of processes unless the value of \deltassm is the same for that set.  If one
wants to make a comparison at the percent level, it has to be done on a mode-by-mode basis, and to do that we need to
build a next generation of experiments to record and analyse ${\cal{O}}(50-100) \iab$ of data.  The two proposed
experiments SuperB and SuperKEKB will be able to make such measurements. If one compares the measured values of
$\sintwobetaeff-\sintwobeta$ for the $b \to d$ processes which have a tree and a loop contribution it is clear that
they are consistent with the SM expectation. At future \B factories it will be possible to extend this approach to
making comparisons of the precision measurements of $\alpha$ and $\gamma$ from different decay channels.

%%%%%%%%%%%%%%%%%%%%%%%%%%%%%%%%%%%%%%%%%%%%%%%%%%%%%%%%%%%%%%%%%%%%%%%%%%%%%%%%%%%%%%%%%%%%%%%%%%%%%%%%%%%
\subsection{Side measurements}
\label{sec:sidemeasurements}
%%%%%%%%%%%%%%%%%%%%%%%%%%%%%%%%%%%%%%%%%%%%%%%%%%%%%%%%%%%%%%%%%%%%%%%%%%%%%%%%%%%%%%%%%%%%%%%%%%%%%%%%%%%

This section discusses the measurements of $\modvub$, $\modvcb$, and $|\vtd/\vts|$ in turn.  All of these quantities
are can be used to constrain the unitarity triangle. The measurements of $\modvub$ and $\modvcb$ use the semi-leptonic
decays $B \to X \ell \nu$ where $X=X_u$ or $X_c$ and it is possible to put constraints on $|\vtd/\vts|$ by measuring
$\B \to X_{d, s}\gamma$ decays.

\subsubsection{Measuring \modvub}
\label{sec:sidemeasurements:vub}

\begin{wrapfigure}{r}{0.6\textwidth}
\centerline{
\includegraphics[width=0.6\textwidth]{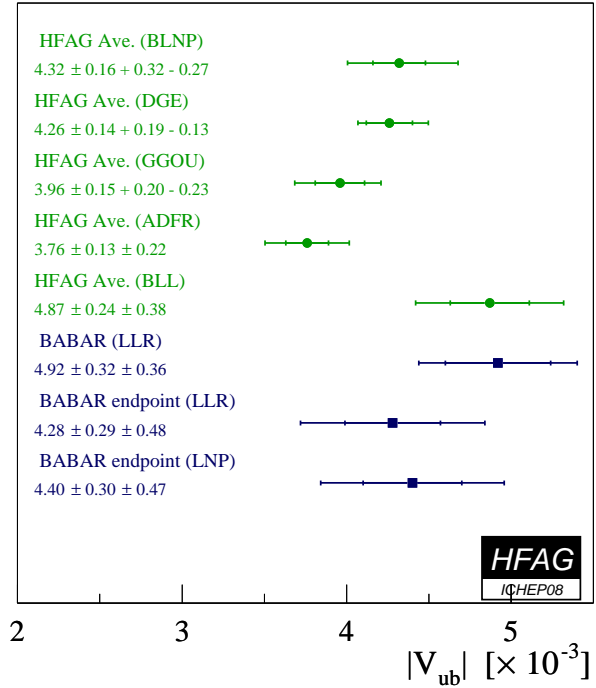}
}
 \caption{Constraints on \modvub compiled by HFAG~\cite{ref:hfag}.}\label{fig:vub}
\end{wrapfigure}
The branching ratios of \B decays to $u \ell \nu$ semi-leptonic final states are proportional to $\modvub$ for a
limited region of phase space. In order to reduce backgrounds in these measurements, both \B mesons in the event are
reconstructed using the so-called {\em recoil} method.  This involves reconstructing the inclusive or exclusive $b\to u
\ell \nu$ signal, as well as reconstructing everything else in the event into a fully reconstructed final state (i.e.
one with no missing energy). If this is done correctly for a \BBb event, then the missing 4-momentum in the centre of
mass will correspond to the 4-momentum of the undetected $\nu$ from the signal decay.  The recoil method results in low
signal efficiencies, typically a few percent, however most of the non-\B background will have been rejected from the
selected sample of events and the signal sample is relatively clean.  Once isolated, it is possible to measure the
partial branching fraction of a decay as a function of a phase space variable, including the $q^2$ of the $\ell\nu$ in
the final state, the invariant mass of the $X_u$, missing mass (corresponding to the neutrino), or energy of the
lepton.

Given the partial branching fraction measurement, theoretical input is required in order to compute $\modvub$.  There
are several schemes available to convert the partial branching fraction to a measurement of \modvub (ADFR, BLNP, BLL,
DGE, GGOU, LLR, and LNP), and all of these schemes ~\cite{ref:vubtheory} give compatible results~\cite{ref:hfag}.
Figure~\ref{fig:vub} shows the different values of \modvub extracted from the data for the different schemes where the
LLR and LNP schemes use $B\to X_u \ell \nu$ decays normalised to $B\to X_s \gamma$ decays in order to determine
\modvub.

\subsubsection{Measuring \modvcb}
\label{sec:sidemeasurements:vcb}

The recoil method discussed above is also used in order to isolate signals in the measurement of \modvcb. Only two
decay channels are considered (i) $B \to D\ell^+ \overline{\nu}$ and (ii) $B^+ \to D^{*0}\ell^+ \overline{\nu}$ where
the partial branching fraction of these decays is proportional to \modvcb up to some form factor.

The partial branching fraction of $B \to D\ell^+ \overline{\nu}$ is proportional to $G^2\modvcb^2$, where $G$ is a form
factor that depends on kinematic quantities.  As the measurement is statistically limited, seven (nine) different $D^0$
($D^+$) daughter decays into final states with neutral and charged pions and kaons are reconstructed. The results
obtained using a combined fit to all data are $G(1)\modvcb = (43.0 \pm 1.9 \pm 1.4)\times 10^{-3}$, where $\modvcb =
(39.8\pm 1.8\pm 1.3 \pm 0.9)\times 10^{-3}$~\cite{ref:hfag} where this result is dominated by
\babar~\cite{ref:babarvcba,ref:babarvcbb}. Errors are statistical, systematic and from the form factor dependence.
Figure~\ref{fig:vcb} shows the distribution of $G(1)\modvcb$ versus $\rho^2$ obtained, where the form factor $G$
depends on the shape parameter $\rho^2$.

The partial branching fraction of $B^+ \to D^{*0}\ell^+ \overline{\nu}$ is proportional to $F^2 \modvcb^2$, where $F$
is a form factor that depends on kinematic quantities.  The measurement of \modvub using this mode is systematically
limited, and as a result the only $D^{*0}$ daughter decay channel considered is to a $D^0 \pi$ final state, where the
$D$ meson subsequently decays to $K^+\pi^-$.  The values of $F(1) \modvcb$ and the slope parameter $\rho^2$ are
extracted from a three dimensional fit to data, where the discriminating variables in the fit are the mass difference
between the reconstructed $D^*$ and $D$ meson masses $\Delta m$, the angle between the $B$ and the $Y=D^*\ell$ in the
centre of mass $\theta_{BY}^*$ and an estimator for the dot product of the four velocities of the $B$ and the $D^*$.
The results obtained are $F(1)\modvcb = (35.97\pm 0.53)\times10^{-3}$ and $\modvcb = (38.7\pm 0.6 \pm 0.9)\times
10^{-3}$, where the first uncertainty is experimental and the second is theoretical~\cite{ref:hfag} where this result
is dominated by \belle and \babar~\cite{ref:babarvcbb,ref:bellevcb,ref:babarvcbc,ref:babarvcbd}. Figure~\ref{fig:vcb}
shows the distribution of $F(1)\modvcb$ versus $\rho^2$.

\begin{figure}[!ht]
\begin{center}
\resizebox{15.0cm}{!}{
 \includegraphics{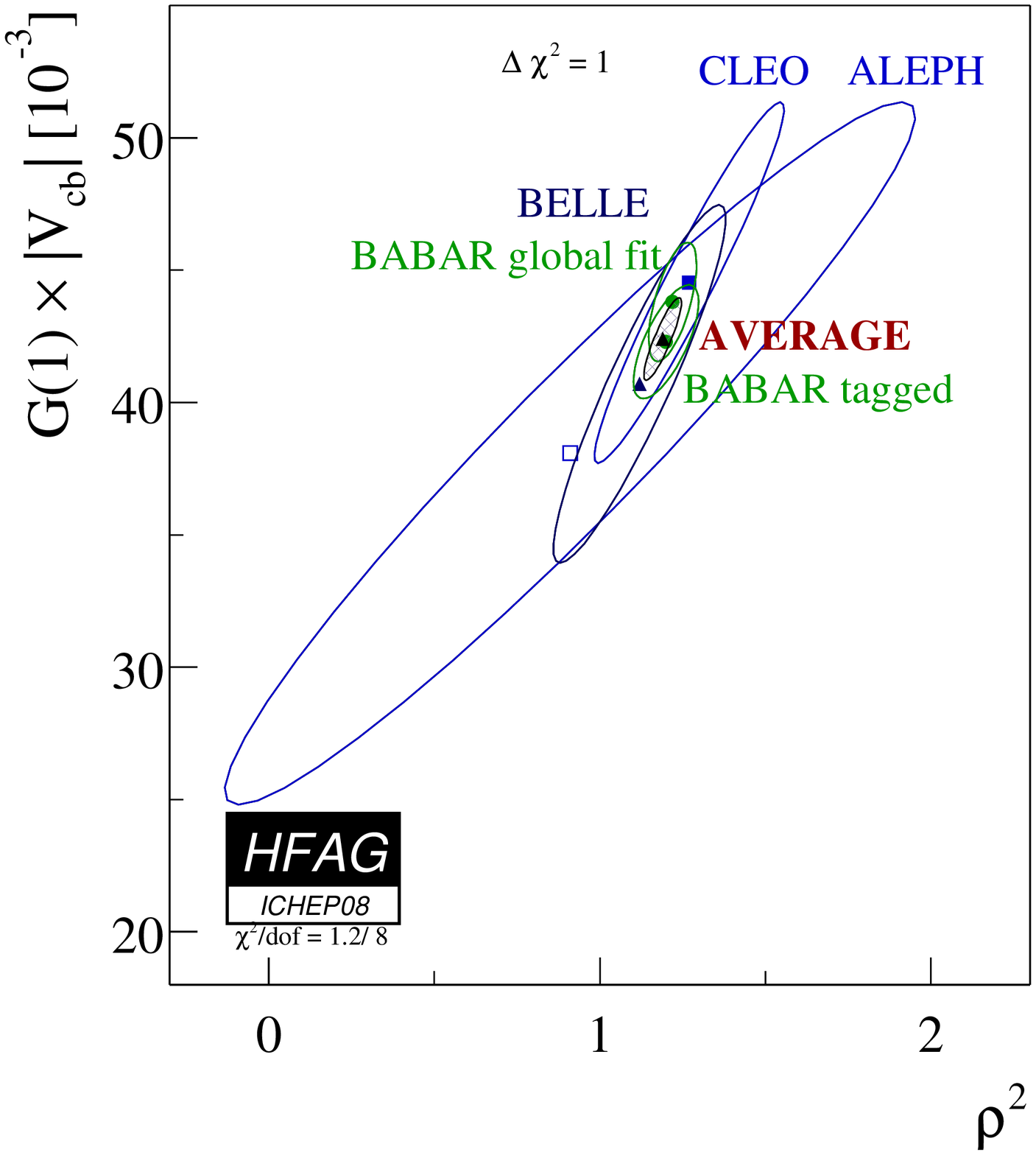}
 \includegraphics{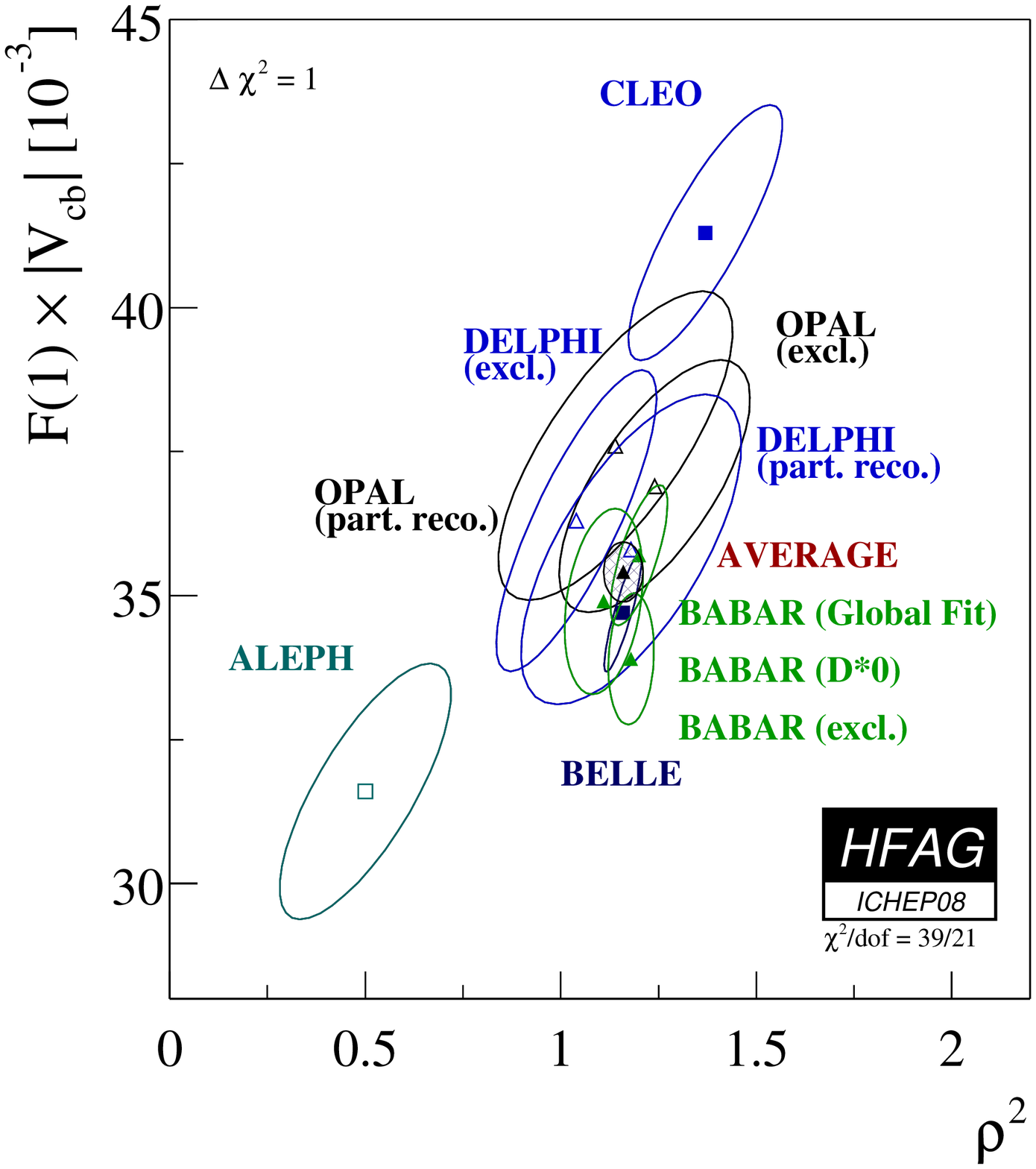}
}
\end{center}
 \caption{Measurements of (left) $G(1)\modvcb$ and (right) $F(1)\modvcb$ versus the slope $\rho^2$ obtained from $D^{(*)}\ell \overline{\nu}$ decays.  These plots are from Ref.~\cite{ref:hfag}.}\label{fig:vcb}
\end{figure}

\subsubsection{Measuring $|\vtd/\vts|$}
\label{sec:sidemeasurements:vcb}

It is possible to measure the ratio $|\vtd/\vts|$ using $B\to X_d \gamma$ and $B\to X_s \gamma$ decays as outlined by
Ali, Asatrian and Greub~\cite{ref:ali_vtdvts}. The branching fractions of these processes depend on \modvtd and
\modvts, respectively. These are Flavour Changing Neutral Currents that are sensitive to new physics, where the leading
order contributions are electroweak loop amplitudes. \babar perform an inclusive analysis of $B\to X_d \gamma$ decays
where $X_d$ is reconstructed from between two and four $\pi$ mesons, or a $\pi^+\eta$ final state, and extract a
branching fraction in two regions of the invariant mass $m_X$ of $X_d$~\cite{ref:babar:dgamma}.  \belle perform an
exclusive analysis and reconstruct $X_d$ in $\rho$ and $\omega$ final states~\cite{ref:belle:dgamma}.  The branching
fractions measured are summarised in Table~\ref{tbl:dgamma}.

\begin{table}[!ht]
\begin{center}
\begin{tabular}{|l|cc|} \hline Experiment & Region/mode & ${\cal {B}}$ ($\times 10^{-6}$) \\ [2pt]
\hline \noalign{\vskip1pt}
 \babar          & $0.6 < m_X < 1.0$ $GeV/c^2$ & $1.2\pm 0.5 \pm 0.1$ \\ [2pt]
 \babar          & $1.0 < m_X < 1.8$ $GeV/c^2$ & $2.7\pm 1.2 \pm 0.4$ \\ [2pt] \hline \noalign{\vskip1pt}
 \belle          & $\B^+\to \rho^+\gamma$ & $0.87^{+0.29}_{-0.27}\,^{+0.09}_{-0.11}$ \\ [2pt]
 \belle          & $\B^0\to \rho^0\gamma$ & $0.78^{+0.17}_{-0.16}\,^{+0.09}_{-0.10}$ \\ [2pt]
 \belle          & $\B^0\to \omega\gamma$ & $0.40^{+0.19}_{-0.17}\,^{+0.09}_{-0.10}$ \\ [2pt] \hline
\end{tabular}
\end{center}
\caption{Branching fraction $({\cal {B}})$ measurements for $B\to X_d \gamma$. Inclusive measurements shown
are from \babar, and exclusive measurements shown are from \belle.\label{tbl:dgamma}}
\end{table}

The constraint on $|\vtd/\vts|$ obtained using these measurements are $0.195^{+0.020}_{-0.019}{\rm(expt)} \pm 0.015
{\rm(theory)}$ and $0.177\pm 0.043{\rm(expt)} \pm 0.001 {\rm(theory)}$ from \belle and \babar, respectively.  The small
theoretical uncertainty on the \babar measurement is the result of the method used to determine $|\vtd/\vts|$ from
data.

%%%%%%%%%%%%%%%%%%%%%%%%%%%%%%%%%%%%%%%%%%%%%%%%%%%%%%%%%%%%%%%%%%%%%%%%%%%%%%%%%%%%%%%%%%%%%%%%%%%%%%%%%%%
\section{Tests of \CPT}
\label{sec:cpt}
%%%%%%%%%%%%%%%%%%%%%%%%%%%%%%%%%%%%%%%%%%%%%%%%%%%%%%%%%%%%%%%%%%%%%%%%%%%%%%%%%%%%%%%%%%%%%%%%%%%%%%%%%%%

The combined symmetry of \C, \P and \T otherwise written as \CPT is conserved in locally gauge invariant quantum field
theory.  The role of \CPT in our understanding of physics is described in more detail in
Refs.~\cite{luders_1954,jost_1957,pauli_1957,dyson_1958} and an observation of \CPT violation would be a sign of new
physics.  \CPT violation could be manifest in neutral meson mixing, so the \B factories are well suited to test this
symmetry. The contribution to these proceedings by Nierst describes the phenomenon of neutral meson mixing in detail in
terms of the complex parameters $p$ and $q$. It is possible to extend the formalism used by Nierst to allow for
possible \CPT violation, and in doing so the heavy and light mass eigenstates of the \Bz meson \BH and \BL become
\begin{equation}
\ket{\BLH} = p \sqrt{1 \mp z} \ket{\Bz} \pm q  \sqrt{1 \pm z} \ket{\Bzb},\nonumber
\end{equation}
where \Bz and \Bzb are the strong eigenstates of the neutral \B meson. If we set $z=0$ we recover the \CPT conserving
solution and if \CP and \CPT are conserved in mixing then $|q|^2+|p|^2=1$.

Two types of analysis have been performed by \babar to test \CPT. The first of these uses the \Bflav sample that
characterises the dilution and resolutions for the Charmonium \sintwobeta analysis discussed in
Section~\ref{sec:cpviolation} along with the Charmonium \CP eigenstates $\Bz \to J/\psi \kz$, $\psi(2S)\ks$, and
$\chi_{1c}\ks$ to extract $z$~\cite{cpt:hadronic:babar}. This analysis also uses control samples of charged \B decays:
$\Bp \to \overline{D}^{(*)0} \pi^+$, $J/\psi K^{(*)+}$, $\psi(2S) K^+$, and $\chi_{1c}K^+$ to obtain
\begin{eqnarray}
\modqovp &=& 1.029 \pm 0.013 \stat \pm 0.011 \syst, \nonumber \\
(\mathrm {Re} \lambdacp / |\lambdacp| ) \mathrm{Re} z &=& 0.014 \pm 0.035 \stat \pm 0.034 \syst, \nonumber \\
\mathrm{Im}z &=& 0.038 \pm 0.029 \stat \pm 0.025 \syst, \nonumber
\end{eqnarray}
which is compatible with no \CP violation in $\Bz-\Bzb$ mixing and \CPT conservation.

The second and more powerful type of analysis uses di-lepton events where both \B mesons in an event decay into an
$X^\mp\ell^\pm \nu$ final state tests \CPT.  Di-lepton events can be grouped by lepton charge into three types: $++$,
$+-$ and $--$ where the numbers of such events $N^{++}$, $N^{+-}$ and $N^{--}$ are related to \Deltagamma and $z$ as a
function of \deltat as described in Ref.~\cite{cpt:dileptonsiderial:babar}. Using these distributions we can construct
two asymmetries: the first is a \T/\CP asymmetry
\begin{eqnarray}
{\cal A}_{\T/\CP} =\frac{P(\Bzb \to \Bz) - P(\Bz \to \Bzb)} {P(\Bzb \to \Bz) + P(\Bz \to \Bzb)}
                  =\frac{N^{++} - N^{--}}{N^{++} - N^{--}}
                  = \frac{1 - \modqovp^4}{1 + \modqovp^4}, \nonumber
\end{eqnarray}
and the second is a \CPT asymmetry
\begin{eqnarray}
{\cal A}_{\CPT}(\deltat)  &=& \frac{N^{+-}(\deltat > 0) - N^{+-}(\deltat < 0)}{N^{+-}(\deltat > 0) + N^{+-}(\deltat <
0) }
                  \simeq 2 \frac{\mathrm{Im} z \sin (\dmd \deltat) - \mathrm{Re} z \sinh \left( \frac{\Deltagamma \deltat }{2}\right)}
                             {\cosh \left( \frac{\Deltagamma \deltat }{2}\right) + \cos(\dmd \deltat)}, \nonumber
\end{eqnarray}
where ${\cal A}_{\CPT}(\deltat)$ is sensitive to $\Deltagamma \times \mathrm{Re} z$.
In the Standard Model ${\cal A}_{\T/\CP}\sim 10^{-3}$ and ${\cal A}_{\CPT} = 0$~\cite{beneke_cpt_2003,ciuchini_cpt_2003}.
\babar measure~\cite{cpt:dilepton:babar}
\begin{eqnarray}
\modqovp - 1  &=& (-0.8 \pm 2.7 \stat \pm 1.9 \syst) \times 10^{-3}, \nonumber \\
\mathrm{Im} z &=& (-13.9 \pm 7.3 \stat \pm 3.2 \syst) \times 10^{-3}, \nonumber \\
\Deltagamma \times \mathrm{Re} z &=& (-7.1 \pm 3.9 \stat \pm 2.0 \syst ) \times 10^{-3}, \nonumber
\end{eqnarray}
which is compatible with no \CP violation in $\Bz-\Bzb$ mixing and \CPT conservation. It is possible to study
variations as a function of sidereal time, where 1 sidereal day is approximately $0.99727$ solar
days~\cite{cpt:siderail:theory} where $z$ depends on the four momentum of the \B candidate.  \babar re-analysed their
data to and find that it is consistent with $z=0$ at 2.8 standard deviations~\cite{cpt:dileptonsiderial:babar}. The
constraint on $z$ is shown in Figure~\ref{fig:cpt}.

\begin{figure}[!ht]
\begin{center}
\resizebox{8.0cm}{!}{
 \includegraphics{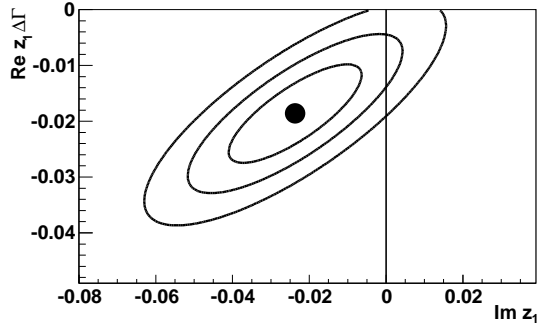}
}
\end{center}
 \caption{Constraints on the imaginary part of $z$ and $\Deltagamma \times \mathrm{Re} z$ using dilepton events at \babar. This figure is reproduced from Ref.~\cite{cpt:dileptonsiderial:babar}.}\label{fig:cpt}
\end{figure}

\section{\B decays to spin one particles}
\label{sec:btovv}

\begin{wrapfigure}{r}{0.5\textwidth}
 \centerline{\includegraphics[width=0.5\textwidth]{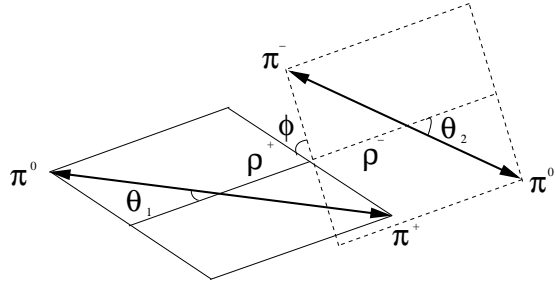}}
 \caption{A schematic of the decay of a \B\ meson via two $\rho$ mesons to a four pion final state.
  The $\rho$ meson final states are shown in their rest frames, and $\phi$ is the angle
  between the decay planes of the $\rho$ mesons.}\label{fig:rhorhoangles}
\end{wrapfigure}

Decays of $B$ mesons to final states with two vector ($J^P=1^-$) or axial-vector ($J^P=1^+$) particles have a number of
interesting kinematic observables that can be used to test theoretical understanding of heavy flavour.  The angular
distribution of such a process where the spin one particles decay into two daughters is a function of three variables:
$\phi$, $\theta_1$ and $\theta_2$,
%\begin{eqnarray}
%\frac{d^3\Gamma}{d\cos\theta_1 d\cos\theta_2 d\phi} &=&
%                  \frac{9}{4}\frac{p}{16 \pi^2 m^2} \times \bigg{\{} \nonumber\\
%                && \frac{1}{4}\sin^2(\theta_1)\sin^2(\theta_2)(|H_{+1}|^2 + |H_{-1}|^2) + \cos^2(\theta_1)\cos^2(\theta_2)|H_0|^2 \nonumber\\
%                &&  +  \frac{1}{2}\sin^2(\theta_1) \sin^2(\theta_2)[\cos(2\phi) Re( H_{+1}H_{-1}^{*} ) - \sin(2\phi) Im( H_{+1}H_{-1}^{*} )] \nonumber\\
%                &&  +  \frac{1}{4}\sin^2(2\theta_1) \sin^2(2\theta_2)[\cos(\phi) Re(H_{+1}H_0^{*}+H_{-1}H_0^{*}) - \sin(\phi) Im(H_{+1}H_0^{*}+H_{-1}H_0^{*}) ]\bigg{\}},\nonumber\\
%\label{eq:angdistributionthreed}
%\end{eqnarray}
where $\phi$ is the angle between the decay planes of the spin one particles, and $\theta_i$ are the angles between the
spin one particle decay daughter momentum and the direction opposite to that of the \Bz in the spin one particle rest
frame. The $\theta_i$ are often referred to as helicity angles. Figure~\ref{fig:rhorhoangles} illustrates these three
angles for the decay $\Bz \to \rho^+\rho^-$.

It is only possible to perform a full angular analysis if we have sufficient data to constrain the unknown observables.
When we search for a rare decay it is normal to perform a simplified angular analysis in terms of the helicity angles,
having first integrated over $\phi$.  On doing this one obtains
\begin{eqnarray}
\frac{d^2\Gamma}{\Gamma d\cos\theta_1 d\cos\theta_2}= \frac{9}{4}\left[f_{\mathrm{L}} \cos^2\theta_1 \cos^2\theta_2 +
\frac{1}{4}(1-f_{\mathrm{L}}) \sin^2\theta_1 \sin^2\theta_2 \right]\label{ eq:angdistributiontwod} \nonumber
\end{eqnarray}
where the parameter $f_{\mathrm{L}}$ is referred to as the fraction of longitudinally polarized events which is given
by
\begin{eqnarray}
\frac{\Gamma_L}{\Gamma} = \frac{|H_0|^2}{|H_0|^2 + |H_{+1}|^2 + |H_{-1}|^2} = f_{\mathrm{L}}, \nonumber
\end{eqnarray}
where the $H_i$ are helicity amplitudes.  It is convenient to analyse the data using the transversity basis when
performing time-dependent \CP studies where the transversity amplitudes are $A_0 = H_0$, $A_\perp=(H_{+1} -
H_{-1})/\sqrt{2}$ and $A_{||}=(H_{+1} + H_{-1})/\sqrt{2}$~\cite{dunietz}. $A_0$ and $A_{||}$ are \CP even and $A_\perp$
is \CP odd. Helicity suppression arguments lead to the expectation of the following hierarchy:
\begin{equation}
A_0 : A_{||} : A_{\perp} \sim {\cal{O}}\left(1\right) : {\cal{O}}\left(\frac{m_R}{m_{B}}\right) :
{\cal{O}}\left(\frac{m_R}{m_{B}}\right)^2, \nonumber
\end{equation}
where $m_R$ is the mass of the spin one resonance and $m_B$ is the $B$ meson mass.  This hierarchy predicts that
$f_{\mathrm L} = 1 - m_R^2/m_B^2$~\cite{ali,kagan,kramer,suzuki}.  The \B factories have measured the fraction of
longitudinally polarized events in a number of different channels, the results of which are shown in
%~\ref{fig:flmeas}.
Table~\ref{tbl:btovv}.

%\begin{wrapfigure}{r}{0.7\textwidth}
% \centerline{\includegraphics[width=0.7\textwidth]{bevan_adrian.flmeas.eps}}
% \caption{Measurements of the $f_{\mathrm{L}}$ that have been performed by the \B factories.  This figure is from
% Ref.~\cite{ref:hfag}}\label{fig:flmeas}
%\end{wrapfigure}

\begin{table}[!ht]
\begin{center}
\begin{tabular}{|l|cc|} \hline Decay Mode & \babar & \belle \\ \hline
 $B^0 \to \phi K^{*0}$ ~\cite{ref:babar:phiksta,ref:belle:phikst} & $0.494\pm 0.34 \pm 0.013$ & $0.45\pm 0.05 \pm 0.02$ \\
 $B^+ \to \phi K^{*+}$ ~\cite{ref:babar:phikstb,ref:belle:phikst} & $0.49\pm 0.05 \pm 0.03$   & $0.52 \pm 0.08 \pm 0.03$ \\
 $B^0 \to \rho^+ \rho^-$ ~\cite{ref:alpharhoprhom:babar,ref:alpharhoprhom:belle} & $0.992\pm 0.024^{+0.026}_{-0.013}$ & $0.941^{+0.034}_{-0.040}\pm 0.030$\\
 $B^0 \to \rho^0 \rho^0$ ~\cite{ref:alpharhozrhoz:babar,ref:alpharhozrhoz:belle} & $0.75^{+0.11}_{-0.14} \pm 0.05$ & \ldots \\
 $B^+ \to \rho^0 \rho^-$ ~\cite{ref:alpharhoprhoz:babar,ref:alpharhoprhoz:belle} & $0.905\pm 0.042^{+0.023}_{-0.027}$ & $0.95 \pm 0.11 \pm 0.02$ \\
 $B^0 \to \omega K^{*0}$ ~\cite{ref:belle:omegakst} & \ldots & $0.56\pm 0.29^{+0.18}_{-0.08}$ \\
 $B^+ \to \omega \rho^{*+}$ ~\cite{ref:babar:omegarho} & $0.82 \pm 0.11 \pm 0.02$ & \ldots \\
 $B^0 \to K^{*0} \overline{K}^{*0}$ ~\cite{ref:kstzkstzbar}& $0.80^{+0.10}_{-0.12}\pm 0.06$ & \ldots \\
 $B^0 \to \rho^0 K^{*0}$ ~\cite{ref:babar:kstrho} & $0.57\pm 0.09 \pm 0.08$& \ldots\\
 $B^+ \to \rho^0 K^{*+}$ ~\cite{ref:babar:kstrho} & $0.96^{+0.04}_{-0.15}\pm 0.05$ & \ldots \\
 $B^+ \to \rho^+ K^{*0}$ ~\cite{ref:babar:kstrho,ref:belle:kstrho} & $0.52 \pm 0.10 \pm 0.04$ & $0.43 \pm 0.11^{+0.05}_{-0.02}$\\
\hline
\end{tabular}
\end{center}
 \caption{Experimental results for $f_{\mathrm {L}}$ from \B meson decays to two vector meson final states.
          The first uncertainty is statistical and the second is systematic.
 \label{tbl:btovv}}
\end{table}

It is clear that the helicity suppression argument works for a number of the decay modes studied. These are all tree
dominated processes such as $\B \to \rho^+\rho^-$.  It is also clear that there are several decay modes that do not
behave in the same way.  The most precisely measured channel that disagrees with the helicity suppression argument is
$\Bz \to \phi K^{*0}$, where $f_{\mathrm{L}} \sim 0.5$.  This discrepancy is often called the `polarization puzzle' in
the literature.  Several papers for example Ref.~\cite{kagan} have highlighted the possibility that new physics could
be the source of the difference, however final state interactions or refined calculations could also account for the
difference.  The decay modes that do not follow the naive helicity suppression argument are all loop dominated.
%Table~\ref{tbl:btovv} summarises the measured values of $f_{\mathrm{L}}$ for the decay modes shown in the Figure.
In addition to studying $B$ decays to final states with two vector particles, it is possible to study vector
axial-vector and two axial-vector final states.  Measurements of these decays could help refine our understanding of
the source of the polarization puzzle.  There are a number of rare decays that have suppressed standard model
topologies, for example $B\to \phi\phi$ and $B \to \phi\rho$.  Experimental limits on these decays are at the level of
a few $10^{-7}$~\cite{babar:phih}. These decays could be significantly enhanced by new physics, and Gronau and Rosner
recently suggested that $\phi-\omega$ mixing could result in a significant enhancement of the $B \to \phi\rho^+$
decay~\cite{gronau_rosner_phirho}.

%%%%%%%%%%%%%%%%%%%%%%%%%%%%%%%%%%%%%%%%%%%%%%%%%%%%%%%%%%%%%%%%%%%%%%%%%%%%%%%%%%%%%%%%%%%%%%%%%%%%%%%%%%%
\section{Summary}
%%%%%%%%%%%%%%%%%%%%%%%%%%%%%%%%%%%%%%%%%%%%%%%%%%%%%%%%%%%%%%%%%%%%%%%%%%%%%%%%%%%%%%%%%%%%%%%%%%%%%%%%%%%

The \B factories have produced a large number of results in many areas of flavour physics.  The ability of the \B
factories to quickly crosscheck each others results has been extremely beneficial to the development of experimental
measurements and techniques over the past decade. Only a small number of these results have been discussed here: those
pertaining to testing the unitarity triangle, \CPT, and \B decays to final states with two spin one particles. These
results are consistent with the CKM theory for \CP violation in the Standard Model. There are a number of measurements
sensitive to new physics contributions that can be made at future experiments such as SuperB in Italy and Super KEK-B
in Japan~\cite{superb,superkekb}.  Such precision tests of flavour physics could be used to elucidate the flavour
structure beyond the Standard Model.

%%%%%%%%%%%%%%%%%%%%%%%%%%%%%%%%%%%%%%%%%%%%%%%%%%%%%%%%%%%%%%%%%%%%%%%%%%%%%%%%%%%%%%%%%%%%%%%%%%%%%%%%%%%
\section{Acknowledgments}
%%%%%%%%%%%%%%%%%%%%%%%%%%%%%%%%%%%%%%%%%%%%%%%%%%%%%%%%%%%%%%%%%%%%%%%%%%%%%%%%%%%%%%%%%%%%%%%%%%%%%%%%%%%

I wish to thank the organizers of this summer school for their hospitality and for giving me the opportunity to discuss
the work described here.  The \B factories have been phenomenally successful and without the collective efforts of both
the \babar and \belle collaborations and the corresponding wealth of theoretical work over the past 45 years I would
not have been in a position to talk on such a vibrant area of research. This work has been funded in part by the STFC
(United Kingdom) and by the U.S. Department of Energy under contract number DE-AC02-76SF00515.

% ****************************************************************************
% BIBLIOGRAPHY AREA
% ****************************************************************************

\begin{footnotesize}
% IF YOU DO NOT USE BIBTEX, USE THE FOLLOWING SAMPLE SCHEME FOR THE REFERENCES
% ----------------------------------------------------------------------------

% ----------------------------------------------------------------------------

\end{footnotesize}
% ****************************************************************************
% END OF BIBLIOGRAPHY AREA
% ****************************************************************************

\end{document}